\documentclass[10pt]{article}

\usepackage[compress]{cite}
\usepackage{amsmath,bm}
\usepackage{hyperref}
\usepackage{enumerate}
\usepackage{bm}
\usepackage{hyperref}
\hypersetup{colorlinks, linkcolor={red!60!black}, citecolor={blue!60!black}, urlcolor={blue!80!black}}
\usepackage{graphicx}
\usepackage[dvipsnames]{xcolor}
\usepackage{amsfonts}
\usepackage{amssymb,amsthm}
\usepackage{mathrsfs}
\usepackage{bbold}

\usepackage[left=2.5cm,right=2.2cm,top=0.5cm,bottom=1.3cm,includeheadfoot]{geometry}

\newcommand{\ham}{\mathcal{H}}
\newcommand{\diff}{\mathcal{D}}
\newcommand{\shift}{{N^x}}
\newcommand{\lapse}{N}
\newcommand{\erad}{E^x{}}
\newcommand{\ephi}{E^\varphi{}}
\newcommand{\krad}{K_x}
\newcommand{\kang}{K_\varphi}

\newcommand{\qbat}{E^x{}}
\newcommand{\qbi}{E^\varphi{}}

\newcommand{\pbat}{K_x}
\newcommand{\pbi}{K_\varphi}

\newcommand{\mass}{M}
\newcommand{\hs}{U}
\newcommand{\s}{{\sigma}}

\renewcommand*{\thefootnote}{\fnsymbol{footnote}}

\numberwithin{equation}{section}

\begin{document}
	\begin{center}
		{\Large\bf 
       {Dynamical theory for spherical black holes in modified gravity}
		}
		\vskip 5mm
		{\large
			Asier Alonso-Bardaji\footnote{e-mail address: {\tt asier.alonso@ehu.eus}}
            and David Brizuela\footnote{e-mail address: {\tt david.brizuela@ehu.eus}}
            }
		\vskip 3mm
		
		{\sl Department of Physics and EHU Quantum Center, University of the Basque Country UPV/EHU,\\
			Barrio Sarriena s/n, 48940 Leioa, Spain}
	\end{center}
 	\vskip 2mm

    \setcounter{footnote}{0}
\renewcommand*{\thefootnote}{\arabic{footnote}}

  		\begin{abstract}
        
        We provide a general algorithm to construct a Hamiltonian, such that its
        dynamical flow covariantly defines any given
        spherically symmetric and static metric. This Hamiltonian
        is defined as a linear combination of the standard (general relativistic) radial diffeomorphism constraint plus
        a Hamiltonian constraint that is appropriately deformed as compared to its corresponding form in general relativity though it does not include higher-derivative terms.
        Therefore, given a static model of spherical gravity, it is possible
        to obtain its Hamiltonian, and, thus, its canonical (second-order) equations of motion.
        A particularly relevant application of this construction is the study
        of regular black holes, where proposed geometries often lack an
        underlying dynamical theory. The present method provides such a theory.  
        In particular, for a wide class of deformations of the Schwarzschild geometry, we explicitly
        obtain their corresponding Hamiltonian. This construction can be further used to covariantly couple matter.
        In this way, one can analyze the backreaction of matter fields on the geometry of interest,
        and, specifically, whether a particular black-hole model
        may emerge as the end state of a dynamical collapse.
\end{abstract}

\section{Introduction}

Regular black holes have long been a topic of interest in gravitational physics, originating from attempts to resolve the singularities present in classical
solutions of general relativity. Some of the earliest models proposed modifications to the Schwarzschild geometry
by introducing additional parameters to avoid the singularity.
Notable examples include the models presented and analyzed in Refs.~\cite{bardeen,poisson:1988wc,Dymnikova:1992ux,Hayward:2005gi,Mars:1996khm,Borde:1996df}.
These constructions were later generalized to incorporate electric charge and angular momentum, giving rise to nonsingular analogs of the Reissner–Nordström and Kerr solutions, as shown, among others, in \cite{Lemos:2011dq, Bambi:2013ufa, Frolov:2016pav}; see also the review \cite{Ansoldi:2008jw}.

In recent years, there has been renewed interest in the subject, driven by advances in quantum gravity and high-energy theory.
More {models} have been proposed, including metrics such as {the one} introduced in Ref.~\cite{Simpson:2018tsi}, which describes a black-to-white hole transition mechanism motivated by proposals for regular gravitational-collapse scenarios, as discussed in Refs.~\cite{Barcelo:2010vc,Barcelo:2014cla, Barcelo:2015uff}. Loop quantum gravity has also inspired several regular black-hole models, including those discussed in Refs.~\cite{Modesto:2008im,Ashtekar:2018lag,Kelly:2020uwj,Alonso-Bardaji:2021yls,Alonso-Bardaji:2022ear,Belfaqih:2024vfk}, which also present a black-to-white hole transition, the Planck star scenario developed in Ref.~\cite{Rovelli:2014cta}, and those where the would-be singularity is unattainable \cite{Boehmer:2007ket,Alesci:2020zfi,Han:2020uhb,Alonso-Bardaji:2024tvp}. For a recent review see Ref.~\cite{Ashtekar:2023cod}. Complementary ideas have emerged from string-theoretical settings \cite{Cano:2018aod, Brandenberger:2021jqs} and from the framework of asymptotic safety \cite{Bonanno:2023rzk}. Several recent reviews
\cite{Lan:2023cvz, Bambi:2023try, Carballo-Rubio:2023mvr, Carballo-Rubio:2025fnc,Carballo-Rubio:2019fnb} provide a broader context, as well as
classifications of regular black holes and their theoretical motivations.

Despite the variety of proposals, constructing a dynamical theory that supports these regular geometries remains an open challenge.
One route is to {introduce an exotic-matter content in} the standard Einstein-Hilbert action. Examples include phantom scalar fields \cite{Bronnikov:2005gm, Bronnikov:2012ch} and Yang-Mills fields \cite{Lavrelashvili:1992ia}. However, these models typically violate energy conditions, casting doubt on their physical viability, though there are attempts to construct regular solutions with matter that obeys the weak energy condition \cite{Ovalle:2023ref}. Another prolific approach involves nonlinear electrodynamics \cite{Ayon-Beato:1998hmi,Balart:2014cga,Ayon-Beato:2004ywd, Dymnikova:2004zc, Bronnikov:2000yz,Ayon-Beato:2000mjt}, which, however, may suffer from generic instabilities \cite{DeFelice:2024seu}.

A different class of approaches leverages modified theories of gravity. These include conformal gravity \cite{narlikar:1977nf,Bambi:2016wdn}, as well as other extensions such as $f(R)$, $f(T)$, and $f(Q)$ gravity \cite{Berej:2006cc,Rodrigues:2015ayd,Junior:2015fya,DAmbrosio:2021zpm,Giacchini:2021pmr,Olmo:2022cui}, often characterized by higher-order derivative or curvature terms. Another notable attempt to provide a dynamical foundation for regular black holes was presented in Ref.~\cite{Ziprick:2010vb}, which draws an analogy between 2-dimensional dilaton gravity and 4-dimensional spherical gravity to construct an action principle for {certain class of modified Schwarzschild} black holes.
In the recent Ref.~\cite{Bueno:2024eig}, the Einstein-Hilbert action is extended by an infinite series of higher-derivative corrections designed to preserve second-order field equations under spherical symmetry. This leads to regular spherical geometries, but this method is only valid for dimensions higher than four. 

While each of these frameworks offers valuable insights, they also come with various challenges, such as {violations of} energy conditions, instabilities, or theoretical ambiguities tied to higher-order dynamics. A different approach, recently presented in Ref.~\cite{Carballo-Rubio:2025ntd},
is based on two-dimensional Horndeski theory and seeks to generalize Einstein equations in a way that encompasses
all possible second-order field equations for spherical metrics. In essence, this is the same philosophy that we
develop in this work. However, we focus on the canonical formulation and, based on our previous results \cite{Alonso-Bardaji:2023vtl},
we provide a novel and systematic method to explicitly obtain a Hamiltonian (quadratic in derivatives), such that its dynamical flow covariantly defines a given static and spherically
symmetric four-dimensional geometry.
In other words, given a spacetime with these features, we build a modified version of the Hamiltonian of general relativity that generates
equations {with up to second-order derivatives and} with the given metric as solution. The formalism is covariant in the sense that the solution to the equations in different gauges will provide the same geometry, though in a different coordinate chart.

This procedure allows for a direct {and covariant (minimal)} coupling of matter fields to the geometry of interest, 
which offers a well-defined dynamical theory grounded in Hamiltonian mechanics.
In particular, it can be used to analyze whether a proposed regular black-hole spacetime represents the endpoint of gravitational collapse. {Therefore, the present method} bridges the gap between static geometries and their potential dynamical origin.

The rest of the article is organized as follows. {Sections~\ref{sec.hamiltonian} and \ref{sec.killing} are a brief review, where} we summarize the main results of Ref.~\cite{Alonso-Bardaji:2023vtl}; namely, the formulation of the covariant Hamiltonian and its associated spherical geometry.
In Section~\ref{sec.reconstruction} we {propose a closed algorithm to solve} the reverse-engineering problem of constructing the Hamiltonian from a given line element. In particular, we provide a closed analytic expression for {the Hamiltonian corresponding to a large class of} modifications of the Schwarzschild spacetime. In Section~\ref{sec.matter}, we present the Hamiltonian for matter fields minimally coupled to any of the above geometries,
and we end up the main text with a brief discussion of our results in Section~\ref{sec.conclusion}.
For illustration, in Appendix~\ref{app.gauges} we explicitly solve the general equations
of motion in two different gauges that define the static and homogeneous charts, respectively.
In Appendix~\ref{app.examples}, we display the Hamiltonian constraint for several metrics of interest, which
include black holes in general relativity, the Bardeen~\cite{bardeen}, Hayward~\cite{Hayward:2005gi}, and Simpson-Visser~\cite{Simpson:2018tsi} regular black holes, as well as several geometries motivated by loop quantum gravity \cite{Kelly:2020uwj,Alonso-Bardaji:2021yls,Alonso-Bardaji:2022ear,Belfaqih:2024vfk}.
Finally, some technical computations are presented in Appendices \ref{app.checkhs} and \ref{app.extra}.

\section{The phase-space description}\label{sec.hamiltonian}

Let us introduce a phase space designed to covariantly\footnote{By covariant we mean
in a gauge-independent way, so that different gauge choices define the same geometry, though in a different coordinate chart.} describe a spherically symmetric geometry through its dynamical flow.
This phase space will be coordinatized by four fields: $\krad, \erad, K_\varphi,$ and $E^\varphi$.
Each of these fields depends on two variables: the time $t$, which parametrizes the dynamical evolution
of the system, and an additional variable $x$, which, as will
be made explicit below, will represent the radial coordinate in the spherically symmetric manifold.
We choose the symplectic structure to be canonical, such that, for a given $t$,
the only nonvanishing Poisson brackets read
\begin{align}
\{\krad(t, x_1),\erad(t, x_2)\}&=\delta(x_1-x_2),\\
\{\kang(t, x_1),\ephi(t, x_2)\}&=\delta(x_1-x_2).
\end{align}
In order to define the Hamiltonian that will generate the dynamical flow,
let us introduce the following two objects, 
\begin{align}
    \label{eq.diff}\diff&:=-\krad\erad'+\ephi\kang',\\
\label{hamSO3vacmod+}
    \ham&:={\mathfrak{g}}\Bigg(\qbi V-\frac{\qbi}{W}\sin^2\big(\omega\pbi\big)\frac{d }{d\qbat}\left(\!\frac{A\,W}{\omega^2}\!\right) +\frac{1}{2}\left(\frac{\qbat''}{\qbi}-\frac{{\qbat'}{\qbi'}}{(\qbi)^2}
    +\frac{(\qbat')^2}{2\qbi W}\frac{d W}{d\qbat}\right)\cos^2\big(\omega\pbi+\varphi\big) \nonumber\\[6pt]
    & -\left({\omega}{\pbat}+{\pbi\qbi}\frac{d \omega}{d\qbat} \right)\frac{A}{\omega^2}\sin\big(2\omega\pbi\big)
    -\left(\omega{\pbat}+\pbi\qbi\frac{d\omega}{d\qbat}+\qbi\frac{d\varphi}{d\qbat}\right)\left(\frac{\qbat'}{2\qbi}\right)^{\!2}\sin\big(2(\omega\pbi+\varphi)\big) \Bigg),
\end{align}
where ${\mathfrak{g}}={\mathfrak{g}}(E^x)$, $\omega=\omega(E^x)$, $\varphi=\varphi(E^x)$, $A=A(E^x)$, $V=V(E^x)$, and $W=W(E^x)$ are free functions of the variable $E^x$,
and the prime stands for a derivative with respect to $x$.

Taking their smeared form, $D[s]:=\int s\diff dx$ and $H[s]:=\int s\ham dx$,
it can be shown by direct computation that {the above objects} obey the algebra,
\begin{subequations}\label{eq.hdageneral}
  \begin{align}
  \label{eq.ddqp}  \big\{D[s_1],D[s_2]\big\}&=D\big[s_1s_2'-s_1's_2\big],\\
 \label{eq.dhqp} \big\{D[s_1],{H}[s_2]\big\}&={H}\big[s_1s_2'\big],\\
  \label{eq.hhqp}   \big\{{H}[s_1],{H}[s_2]\big\}&=D\left[
    (s_1s_2'-s_1's_2)F\right],
  \end{align}
\end{subequations}
with the structure function
\begin{align}\label{eq.F}
    F&:=\frac{F_s}{(\qbi)^2},\quad\mathrm{where}\quad
 F_s:={\mathfrak g}^2\cos(\omega K_\varphi+\varphi)\left(
 A\cos(\omega K_\varphi-\varphi)+\left(\frac{{E^x}'}{2 E^\varphi}\right)^2\omega^2\cos(\omega K_\varphi+\varphi)
 \right).
\end{align}
By defining the Hamiltonian of the system as $H[N]+D[N^x]$,
with the smearing functions $N=N(t,x)$ and $N^x=N^x(t,x)$, the dynamical flow
of the system is given by,
\begin{align}\label{eomEx}
\dot{E}^x=\{E^x,H[N]+D[N^x]\},\\
\dot{K}_x=\{K_x,H[N]+D[N^x]\},\\
\dot{E}^\varphi=\{E^\varphi,H[N]+D[N^x]\},\\
\dot{K}_\varphi=\{K_\varphi,H[N]+D[N^x]\},
\end{align}
where the dot stands for a derivative with respect to $t$.
The smearing functions $N$ and $N^x$ are assumed to be nondynamical in the sense
that their corresponding canonical conjugate momenta vanish.
This implies that, in addition to the four evolution equations
above, on shell, i.e., when evaluated on solutions at any
value of $t$, the above expressions {are constrained to vanish},
\begin{align}\label{constraint1}
{\cal D}&=0,\\\label{constraint2}
{\cal H}&=0.
\end{align}
Since the Poisson brackets between them \eqref{eq.hdageneral} close, both ${\cal D}$ and ${\cal H}$
are first-class constraints. And, as such, they generate gauge transformations on phase space.
In terms of degrees of freedom, each first-class constraint removes one degree of
freedom, i.e., a pair of canonical variables. Consequently, the theory possesses
no local propagating degrees of freedom. This means that, when the system is integrated,
the solution is entirely determined by a finite number of integration constants. 

As commented above, the dynamical flow defined
on the phase space will determine a spherically symmetric geometry on a four-dimensional manifold ${\cal M}$
with topology ${\cal M}^2\times {\cal S}^2$, with ${\cal M}^2$ being a two-dimensional
manifold and ${\cal S}^2$ the two-sphere.
More precisely, as explained in Ref.~\cite{Alonso-Bardaji:2023vtl}, the solution to the equations
of motion \eqref{eomEx}--\eqref{constraint2} covariantly defines the spherically symmetric metric
\begin{equation}\label{eq.metrictx}
ds^2=\s N^2 dt^2+\frac{1}{|F|} (dx+N^x dt)^2 +r^2d\Omega^2,
\end{equation}
on ${\cal M}$,
where $F$ is the structure function \eqref{eq.F} evaluated on a solution
of the system \eqref{eomEx}--\eqref{constraint2}, $\s:=-\mathrm{sgn}(F)$ is the signature of the metric,
$d\Omega^2$ is the metric of the two-sphere, and $r=r(E^x(t,x))$ is a scalar function\footnote{Due to the symmetry reduction, this scalar $r$ does not appear in the phase space,
and, contrary to the other metric functions, it is not fixed by the solution of the system \eqref{eomEx}--\eqref{constraint2}.
In this sense, one is free to {choose $r$ as} any function of $E^x$, since $E^x$ is the only phase-space variable that is a spacetime scalar.
If, instead {of the symmetry reduced version \eqref{eq.hdageneral},} one had at hand the full hypersurface deformation algebra, one would be able to read all the metric functions from there.}
on the manifold ${\cal M}^2$. 
Therefore, $(t,x)$ correspond
to any generic coordinates on ${\cal M}^2$. Solving the equations of motion \eqref{eomEx}--\eqref{constraint2}
in a given gauge, which in particular implies a choice of lapse $N$ and shift $N^x$,
one obtains the form of the metric in a certain coordinate system.
Therefore, the gauge freedom on the phase space corresponds to the diffeomorphism
invariance on the manifold.
This is why the algebra \eqref{eq.hdageneral} is referred to as the hypersurface deformation algebra.
More precisely, it is possible to show that the gauge transformations generated by ${\cal D}$
on the phase space correspond to deformations on a hypersurface of constant $t$
on the manifold. Therefore, ${\cal D}$ is the diffeomorphism constraint. In turn,
the gauge transformations generated by ${\cal H}$ {lead to deformations in the} normal {direction}
to the hypersurfaces of constant $t$, and thus we will call ${\cal H}$ the Hamiltonian constraint.

In fact, since it depends on several free functions, the expression \eqref{hamSO3vacmod+} can be considered as a family of Hamiltonian constraints.
Its specific form \eqref{hamSO3vacmod+} was the main result of Ref.~\cite{Alonso-Bardaji:2023vtl} 
and it is the most general Hamiltonian constraint quadratic in derivatives of the phase-space variables, such that
it forms a closed hypersurface deformation algebra \eqref{eq.hdageneral} with the diffeomorphism
constraint \eqref{eq.diff}, and thus its dynamical flow covariantly defines the metric \eqref{eq.metrictx} on the manifold ${\cal M}$.
Its form is unique up to canonical transformations that leave invariant the diffeomorphism constraint
and do not include derivative terms \cite{Alonso-Bardaji:2023vtl}. An equivalent result was obtained in Ref. \cite{Bojowald:2023xat},
as shown in Appendix~B of Ref. \cite{Alonso-Bardaji:2023vtl}.

Here, some comments are in order about the form of the Hamiltonian constraint \eqref{hamSO3vacmod+}.
First, we note that, as long as the complete expression is real,
the arguments of the trigonometric functions may be complex,
changing them to hyperbolic functions. 
{Second, the case $\omega=0$ should be understood as the case of constant $\omega=\lambda$ with the limit $\lambda\to0$,}
which provides a finite and well defined form.
Third, a particular case of this family of Hamiltonian constraints is the one corresponding to vacuum
general relativity {[see Eq.~\eqref{eq.schwarzschild} below]}. This case is given by the following form of the free functions: 
$\mathfrak{g}=\sqrt{\erad}$, $V=-1/(2\erad)$, $A=1$, $W=\sqrt{\erad}$, $\omega=0$, and $\varphi=0$.
With these functions, one can obtain the solution to the equations of motion \eqref{eomEx}--\eqref{constraint2}
for a given gauge choice and will find that the line element \eqref{eq.metrictx}
corresponds to a certain chart of the Schwarzschild spacetime.

Before ending this section, let us introduce 
the phase-space function \cite{Alonso-Bardaji:2023vtl},
\begin{align}\label{eq.defsm&G}
    \mass&:=\left(\frac{\qbat'}{2\qbi}\right)^2\cos^2\big(\omega\pbi+\varphi\big)-A\frac{\sin^2\big(\omega\pbi\big)}{\omega^2}.
\end{align}
On the one hand, this function allows us to express the Hamiltonian constraint in the following compact form,
\begin{align*}
\ham  &={\mathfrak{g}}\frac{\qbi}{\qbat'}\Bigg(\!\!\left( {V}\!+\frac{M}{W}\frac{d W}{d\qbat}\right)\qbat'+\mass'
     -\frac{\diff}{\qbi}\frac{\partial \mass}{\partial\pbi} 
    \Bigg).
\end{align*}
This expression is valid on all the phase space, even without
evaluating on solutions. However, making use of the equations of motion,
in particular of the constraints \eqref{constraint1}--\eqref{constraint2}, it is immediate
to see that\footnote{
Contrary to what was done in Ref. \cite{Alonso-Bardaji:2023vtl}, where $\approx$
was used to denote on-shell equalities,
here we will not use a specific notation for such equalities.
In fact, when working on the manifold, everything must be understood to be defined on-shell.
}
\footnote{We would like to note that there is a typo in Eq. (59) of Ref.~\cite{Alonso-Bardaji:2023vtl},
and the present Eq. \eqref{monshell} is the correct relation between $M$, $V$, and $W$.},
\begin{align}\label{monshell}
M&=-\frac{1}{W}\int W V d\erad,
\end{align}
and thus, on-shell, i.e., when evaluated on solutions and, in particular on the manifold,
$M$ is a function of $\erad$ only. On the other hand, making use of the
function $M$, we can also write the structure function \eqref{eq.F} as
\begin{align}
    \label{eq.Fs}
F_s&=   {\mathfrak{g}}^2\Big({A}\cos^2(\varphi)+\omega^2\mass\Big).
\end{align}
This expression is valid off shell, but it also shows that, on the manifold,
$F_s$ is a function of $\erad$ only.

\section{Geometric properties}\label{sec.killing}

In order to analyze the geometry \eqref{eq.metrictx},
it is convenient to define the gradient of the area radius function $r$ as the field $v_B:=\nabla_B r$,
where $\nabla_B$ is the covariant derivative associated to the two-dimensional metric on ${\cal M}^2$,
and thus capital Latin indices 
take the values $B=0,1$ with $(x^0,x^1)=(t,x)$.
{Making use of the equation of motion \eqref{eomEx}, the} norm of this vector {can be written as},
\begin{align}\label{eq.hs}
    U:=v^B v_B =  -4 \,\s\, \left(\frac{d r}{d \qbat}\right)^2\mathfrak{g}^2 \mass\Big(A+\omega^2\mass\Big),
\end{align}
where $A$, ${\mathfrak g}$, and $\omega$ are functions that explicitly appear in the Hamiltonian constraint, while
$M$ must be understood as given by Eq.~\eqref{monshell} in terms of $V$ and $W$.

In Ref.~\cite{Alonso-Bardaji:2023vtl}, we proved that, in addition to the Killing vector fields that
define the spherical symmetry, the metric \eqref{eq.metrictx} possesses a Killing field $\xi^B$,
such that its isometry group is $R\times SO(3)$. More precisely,
if $r=r(E^x)$ is an invertible function and $U\neq0$, the Killing vector reads
\begin{equation}
 \xi^B=\exp\left[\int dr \,
 \nabla_{\!C}\left(\frac{v^C}{\hs}\right)  \right] u^B,
\end{equation}
where we have defined $u^B:=\epsilon^{BC}v_C$, with $\epsilon^{BC}$ being
the covariant Levi-Civita tensor on $\mathcal{M}^2$. In particular,
this implies that $\xi^B$ is orthogonal to $v_B$, i.e., $\xi^Bv_B=0$.
The norm of {this} Killing field reads
\begin{align}\label{killingmodule}
 G:=\xi^B\xi_ B
 =-\frac{W^2\hs}{4\mathfrak{g}^2F_s}\left(\frac{dr}{dE^x}\right)^{-2}.
\end{align}
Note that, as usual, the Killing field is defined up to a global nonphysical constant $c$.
In fact, a rescaling of the Killing $\xi^B\to c\, \xi^B$, along with $W\to c\, W$, leaves the whole system
(the Hamiltonian constraint, the equations of motion, and their corresponding solutions)
invariant.

As explicitly given in Ref.~\cite{Alonso-Bardaji:2023vtl}, whenever $v_B\neq 0$, one can use the frame given by $v_B$ and its orthogonal
 $u_B$ to write the metric as
\begin{align}
ds^2=\frac{1}{U}\big(\sigma u_B u_C +v_B v_C\big)dx^Bdx^C+ r^2 d\Omega,
\end{align}
or, equivalently, in terms of the Killing field $\xi^B$ as
\begin{align}
ds^2 &=\left(\frac{1}{G}\xi_B\xi_C+\frac{1}{U}v_B v_C\right)dx^Bdx^C+ r^2 d\Omega.
\end{align}
In order to express this line element in Schwarzschild coordinates, we just need to choose the $t$ coordinate such
that $\partial_t$ is the Killing vector, i.e., $(\xi^0=1,\xi^1=0)$,
and further set the area-radius function as the $x$ coordinate
$r=x$. In this way, we get
\begin{align}\label{metric.diagonal}
    ds^2 = G dt^2 +\frac{1}{\hs}dr^2 +r^2 d\Omega^2.
\end{align}
Note that for Lorentzian spacetimes ($\s=-1$), $F_s$ is positive, and thus $GU\leq0$.
Hence, the spacetime can be split in nonoverlapping static (where $G<0$
and $U>0$) and homogeneous (where $G>0$ and $U<0$) regions, where the hypersurfaces of constant $r$ are timelike and spacelike, respectively. {See Ref.~\cite{Alonso-Bardaji:2023vtl} for further details}.

The {diagonal} chart \eqref{metric.diagonal} breaks down at some hypersurfaces, such as {at} the Killing horizons, where $G=0$, and it thus must be restricted to
an either static or homogeneous region. However,
the change of coordinates $t=\tau+a(r)$, with 
\begin{align}\label{eq.transf}
    \frac{da}{dr}=\varepsilon\sqrt{\frac{\hs-1}{G\hs}},
\end{align}
for a constant $\varepsilon=\pm 1$,
renders the line element \eqref{metric.diagonal} into the form
\begin{align}
    ds^2&=
    G d\tau^2 +2\varepsilon\sqrt{\frac{G}{\hs}(\hs-1)}d\tau dr +dr^2+r^2d\Omega^2,
\end{align}
which is regular at $G=0$.
This chart covers both homogeneous and static regions, including their boundaries $G=0$,
which are null hypersurfaces.
One can invert the coordinate transformation \eqref{eq.transf}, and regain \eqref{metric.diagonal}.
But, since \eqref{eq.transf} diverges precisely at $G=0$, one must restrict the transformation to one
of the branches with either $G<0$ or $G>0$, which then leads to the static or homogeneous diagonal line element \eqref{metric.diagonal}.

For completeness,
in Appendix~\ref{app.gauges}, we explicitly show that the metric in the {diagonal} chart \eqref{metric.diagonal}
is indeed a solution of the Hamilton equations generated by \eqref{hamSO3vacmod+}.
This is a computation of interest to see how a gauge-fixing condition
on the phase space defines a specific chart in the spacetime manifold. 

\section{Construction of the Hamiltonian constraint
from any spherically symmetric and static metric}\label{sec.reconstruction}

Given a spherically symmetric and static geometry,
the problem under consideration is to obtain its corresponding Hamiltonian
constraint, such that the dynamical flow it generates defines
such geometry.
As we will make explicit in this section,
with the results presented above, this turns out
to be {quite} straightforward. 

As a first step one needs to write the metric of interest in the diagonal form,
\begin{equation}\label{eq.reconschart}
 g(r) dt^2+ \frac{dr^2}{f(r)}+r^2d\Omega,
\end{equation}
with $g<0$ and $f>0$, which is always possible due to the assumption of spherical symmetry and staticity. 
Comparing \eqref{eq.reconschart} with $\eqref{metric.diagonal}$, from here one can read off the functions $G=g(r)$ and $U=f(r)$.
In the next step one simply needs
to choose the free functions $r=r(E^x)$, $A=A(E^x)$, $\omega=\omega(E^x)$, $\varphi=\varphi(E^x)$, $\mathfrak{g}=\mathfrak{g}(\erad)$, $V=V(E^x)$, and
$W=W(E^x)$, 
in such a way that relations \eqref{eq.hs} and \eqref{killingmodule} are obeyed---with $M$ and $F_s$ {defined}
by \eqref{monshell} and \eqref{eq.Fs}, respectively---for the given form of $G$ and $U$.
Since there are seven functions to be set, and only two equations, this
construction is not unique and there are five functions of $E^x$ to be arbitrarily chosen.
Once these functions are fixed, one can replace their form in the Hamiltonian constraint \eqref{hamSO3vacmod+},
which would complete the construction.

Let us present this in a more explicit form.
Since we are considering the Lorentzian spacetime \eqref{eq.reconschart}, we fix the signature to $\sigma=-1$
and, for simplicity, we also choose $r=\sqrt{E^x}$
\footnote{As explained above, the area-radius function $r$ is not fixed by the phase-space dynamics.
In general relativity $r=\sqrt{E^x}$, and, for simplicity, we will also take this choice for the general {case,
which also implies $E^x>0$.}}.
Therefore, the system to be solved is 
\begin{align}
    f&=\frac{\mathfrak{g}^2}{r^2} \mass\Big(A+\omega^2\mass\Big),\label{eqf}\\\label{eqg}
    g&= -\frac{W^2\mass\Big(A+\omega^2\mass\Big)}{{\mathfrak{g}}^2\Big({A}\cos^2(\varphi)+\omega^2\mass\Big)},\\\label{eqV}
    V&=-\frac{1}{2rW}\frac{d(M W)}{dr},
\end{align}
where $f=f(r)>0$ and $g=g(r)<0$ are known functions, while $\mathfrak{g}(r)$, $M(r)$, $A(r)$, $\omega(r)$, $W(r)$, $\varphi(r)$, and $V(r)$ are to be set.
Relations \eqref{eqf}--\eqref{eqg} correspond to the equations $f=U$ and $g=G$, respectively,
with the form of $G$ \eqref{killingmodule} and $U$ \eqref{eq.hs} {in terms} of the free functions [that is, writing $F_s$ as in \eqref{eq.Fs}],
while the last relation \eqref{eqV} is simply equation \eqref{monshell} explicitly solved for the function $V$.
As commented above, {as long as} this system is obeyed, one can fix the free functions as desired and simply
replace them in the Hamiltonian constraint \eqref{hamSO3vacmod+} as functions of $E^x$ (taking into account that $r=\sqrt{E^x}$). Depending on the {specific} form of the {metric} functions $f(r)$ and $g(r)$, different
choices may lead to a very different form of the Hamiltonian, which, in some cases, may be easier. 

For instance, the Schwarzschild geometry corresponds to $f=-g=(1-2m/r)$, with a constant mass parameter $m$. This parameter
is not a fundamental constant of theory, it rather appears as an integration constant when solving the Einstein equations.
Therefore, the Hamiltonian constraint should not depend on $m$. It is easy to see that, by fixing the functions
$\mathfrak{g}=r$, $M=(1-2m/r)$, $A=1$, $\omega=0$, $W=r$, $\varphi=0$, and $V(r)=-1/(2r^2)$, the system \eqref{eqf}--\eqref{eqV}
is automatically obeyed. Replacing this choice of functions in \eqref{hamSO3vacmod+} (with $r=\sqrt{E^x}$)
defines the Hamiltonian constraint of vacuum spherically symmetric general relativity,
\begin{align}\label{eq.schwarzschild}
\ham_{\rm vacuum}^{\rm GR}:=-\frac{\qbi}{2\sqrt{\qbat}} \left(1+\pbi^2\right) +\frac{\sqrt{\qbat}}{2}\left(\frac{\qbat''}{\qbi}-\frac{{\qbat'}{\qbi'}}{(\qbi)^2}+\frac{(\qbat')^2}{4 E^x \qbi}\right) -2\sqrt{\qbat} \pbat  \pbi.
\end{align}

In order to analyze a generalization of the Schwarzschild geometry, let us,
for definiteness, consider the following line element
\begin{align}\label{eq.modschw}
ds^2&=-\left(h_1(r)-\frac{2m}{h_2(r)}\right)dt^2+\frac{1}{h_3(r)}\left(h_1(r)-\frac{2m}{h_2(r)}\right)^{-1}dr^2+r^2d\Omega^2,
\end{align}
{with the shape functions $h_i(r)$, for $i=1,2,3$.}
Thus, the metric functions \eqref{eq.reconschart} read
$g=-(h_1-2m/h_2)$ and $f=(h_1-2m/h_2) h_3$, {with
$m$ being} the mass parameter
{of the Schwarzschild reference geometry.}
The goal now is to obtain a corresponding Hamiltonian
constraint independent of the constant $m$, {for which we will further
assume that none of the shape functions $h_i(r)$ depends on $m$}.

Based on the above construction for Schwarzschild,
we begin by fixing the function $M$ to be minus the norm of the Killing field, that is, $M=-g=(h_1-2m/h_2)$.
Note that $M$ does not explicitly appear in the Hamiltonian
and thus its dependence on $m$ is not directly translated to the constraint.
In fact, one can choose $W=h_2$, so that the function $V$, as given by Eq. \eqref{eqV},
reads,
\begin{equation}
V=-\frac{1}{2rh_2}\frac{d}{dr}\left(h_1 h_2\right),
\end{equation}
and neither $W$ nor $V$ depends on $m$.
The other two remaining equations, \eqref{eqf}--\eqref{eqg}, can now be rewritten as,
\begin{align}
A-\omega^2 g&=\frac{r^2}{{\mathfrak g}^2}h_3,\\
A \cos^2(\varphi)-\omega^2 g&=\frac{r^2}{{\mathfrak g}^4}h_2^2h_3.
\end{align}
Since, in these expressions, the only dependence on $m$ is encoded in $g$, at this
point it is natural to choose $\omega=0$.
For simplicity, one can further fix ${\mathfrak g}=r$, which finally leads to
\begin{align}
A=h_3,\\
\cos\varphi=\frac{h_2}{r}.
\end{align}
In this way, replacing the above choice of functions in \eqref{hamSO3vacmod+}, one gets the Hamiltonian constraint,
\begin{align}\label{eq.hamforexamples}
    \ham_{(h_1,h_2,h_3)}:=\Bigg(&-\frac{E^\varphi}{2 h_2}\frac{d\left(h_1 h_2\right)}{dr}-\frac{\pbi^2 E^\varphi}{2 h_2}\frac{d (h_2 h_3)}{dr}
    +\frac{h_2^2}{2\sqrt{E^x}}\left(\frac{\qbat''}{\ephi}-\frac{{\qbat'}{\qbi'}}{(\qbi)^2}\right)
    \nonumber\\&-\left(\frac{2h_2}{\sqrt{\erad}}-3\frac{dh_2}{dr}\right)\frac{h_2}{\erad}\frac{(\qbat')^2}{8\qbi}-2 \sqrt{E^x} h_3\pbat\pbi\Bigg)\Bigg|_{r=\sqrt{E^x}},
\end{align}
where, as stated explicitly, in all instances $r$ should be written as $r=\sqrt{E^x}$,
so that the constraint is a function of the phase-space variables $(E^x,E^\varphi,K_x,K_\varphi)$ and their derivatives.
The form \eqref{eq.modschw} is very general and includes many geometries of interest,
from the usual Reissner-Nordstr{\"o}m-de Sitter black hole, to more general models considered
in the literature of regular black holes \cite{bardeen,Ayon-Beato:1998hmi,Hayward:2005gi,Dymnikova:1992ux,Simpson:2018tsi}.
In Appendix~\ref{app.examples} we present the Hamiltonian \eqref{eq.hamforexamples} for these specific examples.

As can be seen, the shape function $h_2$ appears in a quite intricate and nonlinear form in \eqref{eq.hamforexamples},
and it is thus difficult to interpret the modifications it {yields} on the Hamiltonian.
However, $h_1$ and $h_3$ only appear in one and two terms, respectively, and always with a linear dependence.
In particular, if the shape function $h_2$ takes {the same value as} in Schwarzschild, i.e., $h_2=r$, while $h_1$
and $h_3$ {remain} generic, the Hamiltonian constraint \eqref{eq.hamforexamples} can be written
as
\begin{equation}\label{eq.Hwithh2r}
{\cal H}_{(h_1,h_2=r,h_3)}
={\cal H}_{\rm vacuum}^{\rm GR}+2 \sqrt{E^x} (1-h_3)\pbat\pbi+\frac{E^\varphi}{2\,\sqrt{E^x}}
\left(\frac{d}{dr} \big[r(1-h_1)\big]+\pbi^2\frac{d}{dr} \big[r(1-h_3)\big] \right)\Bigg|_{r=\sqrt{E^x}},
\end{equation}
with ${\cal H}_{\rm vacuum}^{\rm GR}$ being the Hamiltonian constraint of general relativity \eqref{eq.schwarzschild}
that defines the Schwarzschild geometry.
Therefore, the shape functions $h_1$ and $h_3$ can be understood as {generating} some additive
corrections to the constraint of general relativity. In particular, {the term depending on} 
$h_1$ appears coupled to
$E^\varphi$ in a very similar way {as} a minimally coupled scalar field [see {Eq.~\eqref{eq.scalar}} below].

In summary, the dynamical flow generated on the phase space by the Hamiltonian $H_{T}:=\int (N\ham_{(h_1,h_2,h_3)}+N^x\diff)dx$
{covariantly} defines the deformed Schwarzschild geometry \eqref{eq.modschw} for any given form of the shape functions
$h_i(r)$.
More precisely, as explicitly shown in Appendix~\ref{app.checkhs}, 
solving the system of equations
\begin{align}\label{eq.dotexhs}
\dot{E}^x &=\{E^x,H_{T}\},\\
\dot{K}_x &=\{K_x,H_{T}\},\label{eq.dotkxhs}\\
\dot{E}^\varphi &=\{E^\varphi,H_{T}\},\label{eq.dotephihs}\\
\dot{K}_\varphi &=\{K_\varphi,H_{T}\}\label{eq.dotkphihs},\\
\diff&=0,\label{eq.diffhsvanish}\\
\ham_{(h_1,h_2,h_3)}&=0,
\label{eq.hamhsvanish}
\end{align}
with the {gauge choice} $\erad=x^2$ and $\kang=0$, and replacing
the solution
in \eqref{eq.metrictx} leads to the line element \eqref{eq.modschw}, where one should take into account that in this case
$F=h_2^2 h_3/(E^\varphi)^2$ and $r=\sqrt{\erad}$.
Since the construction is covariant,
the solution in any gauge
defines the same geometry. That is, for any gauge choice,
replacing the solution to the above system in \eqref{eq.metrictx}
defines a line element related to \eqref{eq.modschw}
by a coordinate transformation.

\section{Matter coupling}\label{sec.matter}

As already introduced in Ref.~\cite{Alonso-Bardaji:2023vtl}, once we have the covariant Hamiltonian associated to a given geometry, one can minimally couple matter fields.
The total Hamiltonian of the system will read,
\begin{align}
    H_{\rm total}=\int  \Big[(\diff+\diff_m)\shift + (\ham+\ham_m)\lapse\Big]dx,
\end{align}
where $\diff$ and $\ham$ are the gravitational diffeomorphism \eqref{eq.diff} and Hamiltonian \eqref{hamSO3vacmod+} constraint
respectively, while $\diff_m$ and $\ham_m$ correspond to the matter contributions.

Let us write explicitly the contributions corresponding to the specific cases of spherical dust and a scalar field.
Both these cases can be described by an additional pair of conjugate variables $\{\phi(t,x_1),P_\phi(t,x_2)\}=\delta(x_1-x_2)$,
and their contribution to the diffeomorphism constraint reads
\begin{align}
    \diff_m:=\phi' P_\phi.
\end{align}
However,
their addition to the Hamiltonian constraint differs (see Eq.~(102) and Eq.~(106) in Ref.~\cite{Alonso-Bardaji:2023vtl}).
More precisely, for a scalar field $\phi$ with a potential ${\cal V}(\phi)$, it reads 
\begin{align}
    \ham_m^{\rm (scalar)}&:=\frac{\sqrt{|F|}}{2}\left(-\s\frac{P_\phi^2}{\erad}+\erad{(\phi')^2}\right)+\frac{\erad}{2\sqrt{|F|}}\mathcal{V}(\phi),
\end{align}
while, for {a dust field $\phi$}, it takes the form
\begin{equation}
    \ham_m^{\rm (dust)} :=P_\phi\sqrt{1+|F|(\phi')^2}.
\end{equation}
In these expressions we have already assumed $r=\sqrt{E^x}$ and $F=F_s/(E^\varphi)^2$ is the structure function as given in Eq.~\eqref{eq.F},
with the corresponding form of the functions $\mathfrak{g}(\erad)$, $A(\erad)$, $\varphi(\erad)$, and $\omega(\erad)$. 
In general, this structure function can be quite complicated. 
Nonetheless, for the particular case of the Hamiltonian constraint~\eqref{eq.hamforexamples} constructed above,
$\mathfrak{g}=r$, $A=1$, $\cos\varphi=h_2/r$, $\omega=0$, and $F_s$ takes the simple form $F_s=h_2^2h_3>0$.
Therefore, in this case,
the matter contributions for the scalar field (with potential $\mathcal{V}$) and for the dust field read
\begin{align}\label{eq.scalar}
    \ham_m^{\rm (scalar)}&:=\left. \frac{\ephi}{2} h_2\sqrt{h_3}\left(\frac{P_\phi^2}{E^x}+{(\phi')^2E^x}\right)+\frac{E^x\,\mathcal{V}(\phi)}{2\ephi h_2\sqrt{h_3}}\,\right|_{r=\sqrt{E^x}},\\
    \ham_m^{\rm (dust)}&:=\left. P_\phi\sqrt{1+h_2^2h_3\left(\frac{\phi'}{\ephi}\right)^2}\,\,\right|_{r=\sqrt{E^x}},
\end{align}
respectively, and in the arguments of the shape functions, $h_2(r)$ and $h_3(r)$, $r$ must be replaced by $r=\sqrt{\erad}$.

\section{Discussion}\label{sec.conclusion}

Building on our previous results presented in Ref.~\cite{Alonso-Bardaji:2023vtl}, we have formulated and solved the
inverse problem of analytically constructing a Hamiltonian, such that its dynamical flow covariantly
reproduces a given static and spherically symmetric metric \eqref{eq.reconschart}. The Hamiltonian
is expressed as a linear combination of the radial diffeomorphism constraint \eqref{eq.diff}
and the Hamiltonian constraint \eqref{hamSO3vacmod+}, where the latter involves six free functions
constrained by the system \eqref{eqf}–\eqref{eqV}.

Therefore, given any geometry of the form \eqref{eq.reconschart}, one simply
needs to appropriately choose the free functions in \eqref{hamSO3vacmod+},
so that Eqs.~\eqref{eqf}--\eqref{eqV} are obeyed. This choice is generally not unique, as there are more free functions
than equations, and, depending on the specific form of the metric functions $f$ and $g$,
different choices may lead to a simpler expression for the Hamiltonian and its corresponding equations of motion.
However, it is important to emphasize that, since \eqref{eqf}–\eqref{eqg} are purely algebraic and \eqref{eqV}, though differential,
can be interpreted as a definition of the function $V$, the algorithm is fully constructive
and requires no integration.

In particular, we have considered a wide class of deformations of the Schwarzschild geometry \eqref{eq.modschw},
pa\-rame\-trized in terms of three shape functions $h_1(r)$, $h_2(r)$, and $h_3(r)$.
This parametrization is very general and it includes many models of regular black-hole geometries already present in the literature.
For such deformed geometries, we have explicitly obtained a corresponding Hamiltonian constraint \eqref{eq.hamforexamples}.
As commented above, this construction is covariant, and thus, for any gauge choice,
replacing the solution to the Hamilton equations \eqref{eq.dotexhs}--\eqref{eq.hamhsvanish}
in \eqref{eq.metrictx} defines a line element diffeomorphic to \eqref{eq.modschw}.

The shape functions generate different modifications to the Hamiltonian constraint of general relativity, which are in general
difficult to interpret. However, for the case in which the shape function $h_2(r)$ takes the same value as in Schwarzschild,
{i.e., $h_2(r)=r$,}
the Hamiltonian constraint can be written in a compact form: It is just the constraint of vacuum general relativity
\eqref{eq.schwarzschild} plus some additive
correction terms, as shown in Eq.~\eqref{eq.Hwithh2r}.

In the literature on regular black holes, proposed metrics are often introduced
without an underlying theory. In this context, the present method is of particular relevance, as it provides a systematic algorithm to construct such dynamical theory.
This, in turn, allows for a consistent and covariant coupling of matter fields,
providing the full set of Hamilton equations
for the coupled matter and geometric degrees of freedom. As a result, one can study
not only the evolution of matter as described by test fields on the spacetime of interest,
but also analyze its backreaction on the geometry.
Therefore,
this opens up the possibility of testing whether existing regular black-hole geometries
can emerge from a dynamical collapse.
For illustration, the Hamiltonians for some relevant examples of black-hole geometries
(including the Bardeen \cite{bardeen}, Hayward \cite{Hayward:2005gi}, and Simpson-Visser \cite{Simpson:2018tsi} metrics) have been
explicitly worked out in Appendix~\ref{app.examples}.

As a neat example, we find in Appendix~\ref{sec.firstcovariantbh} two different Hamiltonians that define the same geometry. The first option shows
trigonometric functions depending on the
curvature components in the Hamiltonian constraint, as shown in Eq.~\eqref{ourh1}.
However, the second option leads to the constraint \eqref{ourh2}, which simply contains an additive correction
term with respect to the Hamiltonian constraint of {vacuum} general relativity. The existence of two Hamiltonians
leading to the same geometry is not surprising, but it is
important to note that, in each case, the Hamiltonian depends on a different fundamental constant. Therefore,
as this example highlights, the construction of the most natural Hamiltonian for a particular model
strongly depends on the definition of the fundamental constants of the underlying theory.

Our results also feature a broader point regarding covariant formulations of modified gravity, such as those explored in the context of loop quantum gravity. Covariance alone does not restrict the space of admissible deformations or effective geometries: One can {indeed} always construct a covariant model around any chosen line element. {For instance, in Appendix~\ref{app.examples} we obtain the covariant Hamiltonian constraint for some
black-hole models that have been derived {in the context of} loop quantum gravity.}
 {Therefore,} the framework developed here provides a complementary perspective,
where the dynamics and matter content will play a central role in determining the exact form
of an effective theory that encodes effects from the full theory
of loop quantum gravity.

\section*{Acknowledgements}
We thank Marc Schneider for interesting discussions.
This work has been supported by the Basque Government Grant
\mbox{IT1628-22} and by the Grant PID2021-123226NB-I00 (funded by
MCIN/AEI/10.13039/501100011033 and by ``ERDF A way of making Europe'').

\appendix

\section{Static and homogeneous charts from phase space}\label{app.gauges}

In this appendix, we explicitly obtain
the line element \eqref{metric.diagonal} as a solution of the Hamilton equations of motion generated by $H_{TOT}:=H[N]+D[N^x]$,
with $D$ and $H$ being the smeared forms of \eqref{eq.diff} and \eqref{hamSO3vacmod+}, respectively. This procedure is general and valid for any (reasonable) choice of the free functions. 
This is possible with no loss of generality because of the existence of the Killing vector field on the Lorentzian sector, which was proven in Ref.~\cite{Alonso-Bardaji:2023vtl} and allows one 
to write the metric 
in terms of the norm of the Killing field \eqref{killingmodule} and the gradient of the area-radius function \eqref{eq.hs},
as explained
in Section~\ref{sec.killing} of the present paper.

If we replace the functions $G$ and $\hs$ by their on-shell values, the diagonal metric \eqref{metric.diagonal}
explicitly reads
\begin{align}\label{eq:static_chart}
    ds^2&=Gdt^2+\frac{1}{\hs}dr^2+r^2d\Omega^2
    =\frac{\s}{\mathfrak{g}^2}\left[\frac{W^2 \mass(A+\omega^2\mass)}{({A}\cos^2(\varphi)+\omega^2\mass)}{dt}^2 -\frac{1}{4\mass(A+\omega^2\mass)} (d\erad)^2\right]+r^2 d\Omega^2,
\end{align}
with $\mathfrak{g}=\mathfrak{g}(\erad)$, $W=W(\erad)$, $M=M(\erad)$, $A=A(\erad)$, $\omega=\omega(\erad)$, $\varphi=\varphi(\erad)$, and $r=r(\erad)$. 

It is a straightforward though lengthy computation to prove that this chart satisfies the Hamilton equations generated by $H_{TOT}:=H[N]+D[N^x]$. We shall do it in two different gauges:
first with a static gauge choice, by assuming time independence of the variables, and then with
a homogeneous gauge choice, by assuming independence of $x$.

\subsection{Static gauge}\label{app.static}

Let us begin by enforcing the time-independent condition 
\begin{equation}
0=\dot{E}^x:=\{\erad,H_{TOT}\}=N^x {E^x}'+\frac{2N\mathfrak{g}}{\omega \cos(\omega K_\varphi+\varphi)}
[( A +\omega^2 M ) \sin(\omega K_\varphi)\cos(\varphi)+\omega^2 M \cos(\omega K_\varphi)\sin(\varphi)],
\end{equation}
which, for a diagonal line element ($N^x=0$), yields
\begin{align}
 K_\varphi=\frac{1}{\omega}\arctan\left[\frac{-\omega^2 M}{A +\omega^2 M } \tan(\varphi)\right],
\end{align}
and thus $K_\varphi$ is given as a function of $\erad$ only. 
Assuming $\erad'\neq0$, one can then solve the diffeomorphism constraint \eqref{eq.diff} to obtain, 
\begin{align}
    \krad&=\frac{\ephi\kang'}{\erad'}.
\end{align}
Finally, one can extract $\ephi$ and $\lapse$ by direct identification of \eqref{eq:static_chart} with \eqref{eq.metrictx} (see also \eqref{eq.F} and \eqref{eq.Fs}):
\begin{align}
    (\ephi)^2&=-\s\frac{A\cos^2(\varphi)+\omega^2\mass}{4\mass(A+\omega^2\mass)},\\
    \lapse^2&=\frac{W^2 \mass(A+\omega^2\mass)}{\mathfrak{g}^2({A}\cos^2(\varphi)+\omega^2\mass)}.
\end{align}
It is now straightforward to check that all the above conditions satisfy
all the equations of motion, i.e.,
\begin{align}
    \dot{E}^x=0,&&
    \dot{E}^\varphi=0,&&
    \dot{K}_\varphi=0,&&
    \dot{K}_x=0,&&
    \ham=0,&&\mathrm{and}&&
    \diff=0,
\end{align}
and the line element \eqref{eq:static_chart} is thus independent of the time coordinate. Besides, since $N^2>0$, in Lorentzian spacetimes $G<0$ in such a gauge.

\subsection{Homogeneous gauge}

In homogeneous regions, the Lorentzian Killing vector field and the hypersurfaces of constant $r$ are spacelike. This means that, as a function on phase space $E^x$ should obey $\erad'=0$ and, in general, $\dot{E}^x\neq0$. We can always complete the gauge fixing by requiring $\ephi'=0$ and $N^x=0$, which in turn shows that $\kang'=0$, $\krad'=0$, and $N'=0$ on the constraint surface \cite{Alonso-Bardaji:2022ear}.

With these gauge-fixing conditions, the Hamiltonian constraint \eqref{hamSO3vacmod+} reduces to
\begin{align}
    \ham&={\mathfrak{g}}\Bigg(\qbi V-\frac{\qbi}{W}\sin^2\big(\omega\pbi\big)\frac{d }{d\qbat}\left(\!\frac{A\,W}{\omega^2}\!\right)  -\left({\omega}{\pbat}+{\pbi\qbi}\frac{d \omega}{d\qbat} \right)\frac{A}{\omega^2}\sin\big(2\omega\pbi\big)
   \Bigg),
\end{align}
and we can solve $\ham=0$ for
\begin{align}
    \krad&=-\frac{\kang\ephi}{\omega}\frac{d\omega}{d\erad}+\frac{\omega\ephi}{A\sin(2\omega\kang)}\left(V-\frac{1}{W}\sin^2\big(\omega\pbi\big)\frac{d }{d\qbat}\left(\!\frac{A\,W}{\omega^2}\!\right)  \right).
\end{align}
By direct inspection of the line elements \eqref{eq.metrictx} and \eqref{eq:static_chart} [using also \eqref{eq.F} and \eqref{eq.Fs}], we find
\begin{align}
    (\ephi)^2&=\s W^2\mass(A+\omega^2\mass)=-\s \frac{W^2 A^2}{4\omega^2}\sin^2(2\omega\kang),\\
    N^2&=\frac{\s}{4\mathfrak{g}^2\mass(A+\omega^2\mass)}=-\frac{\s\omega^2}{\mathfrak{g}^2A^2\sin^2(2\omega\kang)},
\end{align}
where we have used that
\begin{align}
    \mass&=-A\frac{\sin^2\big(\omega\pbi\big)}{\omega^2}.
\end{align}
Note that in Lorentzian spacetimes ($\s=-1$) both $N^2$ and $(\ephi)^2$ are, as expected, positive (even for imaginary or vanishing $\omega$).

Making use of the above results, we can easily check that
\begin{align}
\dot{E}^x:=\{\erad,H_{TOT}\}&=\frac{N\mathfrak{g}A\sin(2\omega\kang)}{\omega }=\pm1,
\end{align}
that is, $\erad$ is actually playing the role of time. 
From now on, we consider the positive sign for $\dot{E}^x$, corresponding to $N>0$. 
We only need to find $\kang$, and this is straightforward from its equation of motion
\begin{align}
\dot{K}_\varphi&:=\{\kang,H_{TOT}\}=\frac{\omega}{A\sin(2\omega\kang)}\left(V-\frac{1}{W}\sin^2\big(\omega\pbi\big)\frac{d }{d\qbat}\left(\!\frac{A\,W}{\omega^2}\!\right)  -{\pbi}\frac{d\omega}{d\qbat} \frac{A}{\omega^2}\sin\big(2\omega\pbi\big)\right)\nonumber\\
&=\frac{\omega}{A\sin(2\omega\kang)}\left(V-\frac{1}{W}\frac{A}{\omega^2}\sin^2\big(\omega\pbi\big)\frac{d W}{d\qbat}-\sin^2\big(\omega\pbi\big)\frac{d }{d\qbat}\left(\!\frac{A}{\omega^2}\!\right)  -{\pbi}\frac{d\omega}{d\qbat} \frac{A}{\omega^2}\sin\big(2\omega\pbi\big)\right),
\end{align}
which can be simplified to
\begin{align}
\frac{d}{dt}\ln\left[\frac{AW}{\omega^2}\sin^2(\omega\kang)\right]=\frac{V\omega^2}{A\sin^2\big(\omega\pbi\big)},
\end{align}
where we have used that $\erad=t$ in the last step. This can be integrated to give
\begin{align}
K_\varphi=\frac{1}{\omega}\arcsin\sqrt{\frac{\omega^2}{AW} \int V W dt}.
\end{align}
Note that in the limit to general relativity $\omega\to0$, $A=1$, $V=-1/(2\erad)$, and $W=\sqrt{\erad}$,
the solution is, as expected for the Schwarzschild interior, $\kang^2=2m/\sqrt{\erad}-1$, with $m$ an integration constant.

With this, all the variables are fixed 
and all the equations of motion \eqref{eomEx}--\eqref{constraint2} are satisfied with $G>0$.

\section{Hamiltonian constraint for specific line elements}\label{app.examples}

In this appendix we present different geometries of interest that describe spherical black holes,
and construct their corresponding Hamiltonian constraint. All of them fit the parametrization \eqref{eq.modschw} of a deformed Schwarzschild black hole.
The only issue is that, in some cases, the shape functions depend on the
mass parameter $m$. Therefore, the procedure proposed in the main text,
which leads to the constraint \eqref{eq.hamforexamples}, would define a Hamiltonian that
depends on $m$. However, in most of the cases, $m$ appears
multiplying another constant parameter, which can simply be redefined to absorb the
dependence on $m$.
The only exception to this fact is the geometry studied in Section~\ref{app.lqgbarmu}, where the mass dependence
cannot be absorbed by a rescaling of another constant; one would actually need
to define two new independent parameters.

Since, in general,
the underlying dynamical theory is unknown, it is not {\it a priori}
clear which is the fundamental constant (or constants) of the model that
should be present in the Hamiltonian. Therefore, in all the cases where the mass dependence
can be absorbed with a simple rescaling of a constant, we will do so and will then simply
apply the general result \eqref{eq.hamforexamples} obtained in the main text.
{However, if one was interested in other alternative parameters appearing in the Hamiltonian, 
a different construction of the Hamiltonian would be more appropriate, as long as the free functions satisfy the system \eqref{eqf}--\eqref{eqV}.
}

More precisely, {Sections \ref{sec.rn}--\ref{sec.simpsonvisser}} present geometries
that implement the additive corrections of the shape function $h_1$ {and $h_3$}
described in \eqref{eq.Hwithh2r}, while Sections~\ref{sec.hayward}--\ref{sec.bardeen} involve
metrics with the nonlinear corrections originating from a nontrivial form of $h_2$.
An example combining both types of modifications can be found in Section~\ref{sec.genbardeen}. 
Finally,
in Secs.~\ref{app.lqgbarmu}--\ref{sec.firstcovariantbh} we present certain geometries constructed
in the context of loop quantum gravity.
We use these as an example to present some alternative derivations of the Hamiltonian.

\subsection{The Reissner-Nordstr{\"o}m-de Sitter spacetime}\label{sec.rn}

Let us check the construction in the main text for the general spherically symmetric black hole in general relativity,
that is, the Reissner-Nordstr{\"o}m-de Sitter metric,
\begin{align}
    ds^2_{\rm RNdS}=-\left(1-\frac{2M}{r}+\frac{Q^2}{r^2}-\frac{\Lambda}{3}r^2\right)dt^2+\left(1-\frac{2M}{r}+\frac{Q^2}{r^2}-\frac{\Lambda}{3}r^2\right)^{-1}dr^2+r^2d\Omega^2,
\end{align}
which describes a spherical vacuum spacetime with electric charge $Q$ and cosmological constant $\Lambda$.
Comparing this with the line element \eqref{eq.modschw}, it is straightforward to see that
it corresponds to the shape functions $h_1=1+Q^2/r^2-\Lambda r^2/3$, $h_2=r$, and $h_3=1$. 
Replacing these values in \eqref{eq.Hwithh2r} and performing the change $r\to\sqrt{\erad}$,
leads to the Hamiltonian constraint,
\begin{align}\label{eq.rnds}
\ham_{\rm RNdS} := \ham_{\rm vacuum}^{\rm GR}+\frac{\qbi}{2\sqrt{\qbat}} \left(\frac{Q^2}{\erad}+\Lambda \erad\right),
\end{align}
{with ${\cal H}_{\rm vacuum}^{\rm GR}$ as defined in \eqref{eq.schwarzschild}.}
As expected, the charge and cosmological constant appear as {additive contributions} to the Hamiltonian
constraint, as they both can be understood as being originated from a minimal coupling of matter fields.

\subsection{The $\bar{\mu}$-loop black hole}\label{sec.lqgmubar}

We now apply the result in the main text to a popular black-hole geometry
in the context of loop quantum gravity \cite{Kelly:2020uwj},
\begin{align}\label{eq.muoriginal}
    ds_{\bar\mu}^2
    =-\left(1-\frac{2m}{r}+\frac{\alpha}{r^4}\right)dt^2+2\sqrt{\frac{2m}{r}-\frac{\alpha}{r^4}}dtdr +dr^2+r^2d\Omega^2,
\end{align}
with {the constant parameter} $\alpha:=4 m^2\gamma^2\Delta$, 
and provide its corresponding covariant Hamiltonian. 
When diagonalized, this metric reads
\begin{align}\label{eq.mulqg}
    ds^2_{\bar\mu}=-\left(1-\frac{2m}{r}+\frac{\alpha}{r^4}\right)dt^2+\left(1-\frac{2m}{r}+\frac{\alpha}{r^4}\right)^{-1}dr^2+r^2d\Omega^2.
\end{align}
 Hence, the shape functions are given by $h_1=1+\alpha/r^4$, $h_2=r$, and $h_3=1$. By replacing these forms
in \eqref{eq.Hwithh2r}, we find the Hamiltonian constraint,
\begin{align}\label{eq.mulqgham}
\ham_{\bar \mu} :=\ham_{\rm vacuum}^{\rm GR}+\frac{3\alpha\ephi}{2({\erad})^{5/2}},
\end{align}
{with ${\cal H}_{\rm vacuum}^{\rm GR}$ given in \eqref{eq.schwarzschild}.}
Therefore, this covariant construction of the Hamiltonian shows that the so-called loop-quantum-gravity effects considered in Ref.~\cite{Kelly:2020uwj}
can be simply understood as a repulsive potential parametrized by a constant $\alpha$.
In order to illustrate that the above Hamiltonian indeed {(covariantly) yields} the geometry described in Ref.~\cite{Kelly:2020uwj}
{for different gauge choices},
in Appendix~\ref{app.extra} we solve the Hamilton equations in two gauges.
As commented above,
if one wants to consider $\gamma^2\Delta$ instead of $\alpha$ to be the fundamental constant of the model, a different construction might be better suited.

\subsection{The Simpson-Visser black hole}\label{sec.simpsonvisser}

In Ref.~\cite{Simpson:2018tsi} Simpson and Visser presented the metric,
\begin{align}
    ds^2_{\rm SV}=-\left(1-\frac{2m}{\sqrt{\tilde{r}^2+a^2}}\right)dt^2+
    \left(1-\frac{2m}{\sqrt{\tilde{r}^2+a^2}}\right)^{-1}d\tilde{r}^2+(\tilde{r}^2+a^2)d\Omega^2,
\end{align}
with $a\neq 0$. Under the coordinate transformation defined by $r=\sqrt{\tilde{r}^2+a^2}$,
this line element takes the form,
\begin{align}
    ds^2_{\rm SV}=-\left(1-\frac{2m}{r}\right)dt^2+
    \left(1-\frac{2m}{r}\right)^{-1} \left(1-\frac{a^2}{r^2}\right)^{-1}dr^2+r^2 d\Omega^2.
\end{align}
Comparing this with \eqref{eq.modschw}, it is now straightforward to identify the
shape functions $h_1=1$, $h_2=r$, and $h_3=(1-a^2/r^2)$.
By replacing these
in \eqref{eq.Hwithh2r}, we finally obtain its corresponding Hamiltonian constraint
\begin{align}\label{eq.rnds}
\ham_{\rm SV} := \ham_{\rm vacuum}^{\rm GR}
+2a^2\frac{\pbat\pbi}{\sqrt{E^x}}-\frac{a^2}{2}\frac{E^\varphi\pbi^2}{(E^x)^{3/2}},
\end{align}
{with ${\cal H}_{\rm vacuum}^{\rm GR}$ defined in \eqref{eq.schwarzschild}.}

\subsection{The Hayward metric}\label{sec.hayward}

The Hayward metric reads \cite{Hayward:2005gi},
\begin{align}
    ds^2_{\rm Hayward}=-\left(1-\frac{2mr^2}{r^3+\beta}\right)dt^2+\left(1-\frac{2mr^2}{r^3+\beta}\right)^{-1}dr^2+r^2d\Omega^2,
\end{align}
{where we have defined the constant $\beta:=2l^2m$ to absorb an $m$.} 
Then, we simply need to identify the shape functions $h_1=1=h_3$ and $h_2=r+\beta/r^2$, and apply them in the general result \eqref{eq.hamforexamples},
\begin{align}
    \ham_{\rm Hayward}:=&-\left(1-\frac{2\beta}{(\erad)^{3/2}}\right)\left(1+\frac{\beta}{(\erad)^{3/2}}\right)^{-1}\frac{E^\varphi}{2\sqrt{\erad}}\left(1+\kang^2\right)
    +\left(1+\frac{\beta}{(\erad)^{3/2}}\right)^2\frac{\sqrt{E^x}}{2}\left(\frac{\qbat''}{\ephi}-\frac{{\qbat'}{\qbi'}}{(\qbi)^2}\right)
    \nonumber\\&+\left(1-\frac{7\beta}{(\erad)^{3/2}}-\frac{8\beta^2}{(\erad)^3}\right)\frac{(\qbat')^2}{8\sqrt{\erad}\qbi}
    -2 \sqrt{E^x}\pbat\pbi.
\end{align}
{As explained above, if, instead of $\beta$, 
one wishes, for instance, to interpret $l$ as the fundamental constant of the model,
other alternative constructions of the Hamiltonian would be better suited.
}

\subsection{The Bardeen black hole}\label{sec.bardeen}

We now turn our attention to the Bardeen black hole \cite{bardeen}, which is one of the first known
regular black-hole geometries proposed in the literature. The Bardeen metric is given by
\begin{align}\label{eq.bardeenm}
    ds^2_{\rm Bardeen}=-\left(1-\frac{2mr^2}{(q^2+r^2)^{3/2}}\right)dt^2+\left(1-\frac{2mr^2}{(q^2+r^2)^{3/2}}\right)^{-1}dr^2+r^2d\Omega^2,
\end{align}
with a nonvanishing constant parameter $q$. The case $q=0$ restores the Schwarzschild geometry.
Comparing this with \eqref{eq.modschw}, we can thus identify $h_1=1=h_3$ and $h_2=(q^2+r^2)^{3/2}/r^2$.
Considering these shape functions in the general Hamiltonian constraint \eqref{eq.hamforexamples} yields
\begin{align}\label{eq.bardeenH}
    \ham_{\rm Bardeen}:=&-\left(1-\frac{2q^2}{\erad}\right)\left(1+\frac{q^2}{\erad}\right)^{-1}\frac{\ephi}{2\sqrt{\erad}}(1+\pbi^2) +\left(1+\frac{q^2}{\erad}\right)^3\frac{\sqrt{\erad}}{2}\left(\frac{\qbat''}{\ephi}-\frac{{\qbat'}{\qbi'}}{(\qbi)^2}
    \right)\nonumber\\
    &+\left(1+\frac{q^2}{\erad}\right)^2\left(1-\frac{8q^2}{\erad}\right)\frac{(\qbat')^2}{8\sqrt{\erad}\qbi} -2\sqrt{\erad}\pbat\pbi.
\end{align}
Note that the corrections, as parametrized by $q$, appear in a highly nontrivial way in this expression and, in particular,
this Hamiltonian cannot be understood as the one corresponding to vacuum general relativity plus certain additive corrections.
Moreover, the corrections in the first term imply that the modifications cannot be written as a finite power series of $q^2$, {as one can also read from the metric \eqref{eq.bardeenm}.}

\subsection{The generalized Bardeen black hole}\label{sec.genbardeen}

Let us now turn to a regular exact black-hole solution in general relativity describing gravity coupled to nonlinear electrodynamics \cite{Ayon-Beato:1998hmi}:
\begin{align}
    ds^2_{\rm NLED}=-\left(1-\frac{2mr^2}{(q^2+r^2)^{3/2}}+\frac{q^2r^2}{(q^2+r^2)^2}\right)dt^2+\left(1-\frac{2mr^2}{(q^2+r^2)^{3/2}}+\frac{q^2r^2}{(q^2+r^2)^2}\right)^{-1}dr^2+r^2d\Omega^2.
\end{align}
Clearly, this is a generalization of the Bardeen geometry, and in our context it corresponds to the shape functions $h_1=1+q^2r^2/(q^2+r^2)^2$, $h_2=(q^2+r^2)^{3/2}/r^2$, and $h_3=1$. This leads to the following Hamiltonian constraint,
\begin{align}
    \ham_{\rm NLED}:=&-\left(1-\frac{q^2}{\erad}-\frac{3q^4}{(\erad)^2}-\frac{2q^6}{(\erad)^3}\right)\left(1+\frac{q^2}{\erad}\right)^{-3}\frac{E^\varphi}{2\sqrt{\erad}}-\left(1-\frac{2q^2}{\erad}\right)\left(1+\frac{q^2}{\erad}\right)^{-1}\frac{\ephi\pbi^2}{2\sqrt{\erad}}
    \nonumber\\&
    +\left(1+\frac{q^2}{\erad}\right)^3\frac{\sqrt{\erad}}{2}\left(\frac{\qbat''}{\ephi}-\frac{{\qbat'}{\qbi'}}{(\qbi)^2}\right)
    +\left(1+\frac{q^2}{\erad}\right)^2\left(1-\frac{8q^2}{\erad}\right)\frac{(\qbat')^2}{8\sqrt{\erad}\qbi}
    -2 \sqrt{E^x}\pbat\pbi,
\end{align}
which differs from the Bardeen one \eqref{eq.bardeenH} only in the potential term (the first term).

\subsection{Covariant {loop} black holes: Scale-dependent holonomies}\label{app.lqgbarmu}

We now present an example where the method {to construct the Hamiltonian} described in the main text is not the most convenient
one, since the shape functions explicitly depend on the mass parameter $m$, and this dependence cannot be absorbed in an additional free constant (we would need to define two new independent parameters).

More precisely, let us consider the line element \cite{Belfaqih:2024vfk},
\begin{align}
    ds^2_{\Delta}&=-\left(1-\frac{2m}{r}\right)dt^2+\bigg(1+\frac{\Delta}{r^2}\left(1-\frac{2m}{r}\right)\bigg)^{-1}\left(1-\frac{2m}{r}\right)^{-1}dr^2+r^2 d\Omega^2,
\end{align}
where $\Delta>0$ and, as compared to \eqref{eq.hamforexamples}, $h_1=1$, $h_2=r$, and $h_3=1-(1-2m/r)\Delta/r^2$. 
Therefore,
the procedure of the main text would lead to a Hamiltonian with explicit dependence on $m$,
which is something we want to avoid. In fact, in order to absorb this dependence, one could
define $\widetilde\Delta:=m\Delta$, but then the two parameters $\Delta$ and $\widetilde\Delta$
would be present in the Hamiltonian. Let us thus perform an alternative construction, which
will define a Hamiltonian only depending on $\Delta$.

Since $h_1$ and $h_2$ do not depend on $m$,
we can begin the procedure in the same form {as in the main text}, by fixing $M=-g$, $W=r$, $V=-1/(2r^2)$,
$\mathfrak{g}=r$, and $\varphi=0$. Now, the system of equations \eqref{eqf}--\eqref{eqg} is reduced to
the condition,
\begin{align*}
    A+\omega^2\left(1-\frac{2m}{r}\right)=1+\frac{\Delta}{r^2}\left(1-\frac{2m}{r}\right),
\end{align*}
which can be clearly satisfied by fixing $A=1$ and $\omega=\sqrt{\Delta}/r$. {Finally,
replacing these choices in \eqref{hamSO3vacmod+} with $r=\sqrt{E^x}$,
we end up with the following} form of the Hamiltonian constraint,
\begin{align}
      \ham_\Delta&:=-\frac{\qbi}{2\sqrt{\erad}}-\frac{3\sqrt{\erad}\qbi}{2\Delta}\sin^2\left(\sqrt{\frac{\Delta}{\erad}}\pbi\right) +\frac{\sqrt{\erad}}{2}\left(\frac{\qbat''}{\qbi}-\frac{{\qbat'}{\qbi'}}{(\qbi)^2}
    +\frac{(\qbat')^2}{4\qbi{\erad}}\right)\cos^2\left(\sqrt{\frac{\Delta}{\erad}}\pbi\right) \nonumber\\
    & -\left(\sqrt{\erad}{\pbat}-\frac{\pbi\qbi}{2\sqrt{\erad}} \right)\left(1+\frac{\Delta}{\erad}\left(\frac{\qbat'}{2\qbi}\right)^{\!2}\right)\sqrt{\frac{\erad}{\Delta}}\sin\left(2\sqrt{\frac{\Delta}{\erad}}\pbi\right).
\end{align}
Note, in particular, that the limit $\Delta\to0$ yields the classical constraint \eqref{eq.schwarzschild}.

\subsection{Covariant loop black holes: Constant polymerization}\label{sec.firstcovariantbh}

This example was the first covariant implementation of corrections {motivated by loop quantum gravity in spherical symmetry},
and the form of its corresponding {covariant} Hamiltonian constraint is known \cite{Alonso-Bardaji:2021yls,Alonso-Bardaji:2022ear}.
However, we will use {this model} to show that
the same geometry {can also} be derived from an alternative (and simpler) Hamiltonian.

{In a diagonal and static chart, the }metric presented in Refs.~\cite{Alonso-Bardaji:2021yls,Alonso-Bardaji:2022ear} reads
\begin{align}\label{eq.ourmetric}
    ds^2_{r_0}&=-\left(1-\frac{2m}{r}\right)dt^2+\bigg(1-\frac{r_0}{r}\bigg)^{-1}\left(1-\frac{2m}{r}\right)^{-1}dr^2+r^2 d\Omega^2,
\end{align}
with $r_0=2m\lambda^2/(1+\lambda^2)$ and $\lambda$ a nonzero constant. The constant $r_0\in\mathbb{R}^+$ corresponds to the global minimum of the area-radius function $r$,
which is attained in the regular hypersurface that replaces the classical singularity.

The Hamiltonian constraint considered in Refs.~\cite{Alonso-Bardaji:2021yls,Alonso-Bardaji:2022ear} can be reproduced from \eqref{hamSO3vacmod+} by simply setting the free functions
to $A=1$, $\varphi=0$, $\omega=\lambda$, $W=\sqrt{\erad}$, $V=-1/(2\erad)$, and $\mathfrak{g}=\sqrt{\erad}/\sqrt{1+\lambda^2}$,
which clearly satisfies the system \eqref{eqf}--\eqref{eqV} with $M=(1-2m/\sqrt{E^x})$ and $r=\sqrt{E^x}$, and it leads to
\begin{align}\label{ourh1}
    \ham_{\lambda}&:=\frac{1}{\sqrt{1+\lambda^2}}\Bigg[-\frac{{\ephi}}{2\sqrt{{\erad}}}\left(1+\frac{\sin^2{{(\lambda {\kang})}}}{{{\lambda^2}}}\right) +\left(\frac{({\erad}')^2}{8\sqrt{{\erad}}{\ephi}}+\frac{\sqrt{{E}^x}}{2}\left(\frac{{E}^x{}''}{{E}^\varphi}-\frac{\erad'\ephi'}{(\ephi)^2}\right)\right)\cos^2{(\lambda {\kang})} \nonumber\\
    &\qquad\qquad\qquad\;-\sqrt{{\erad}}{\krad}\frac{\sin{(2\lambda {\kang})}}{\lambda}\left(1+\left(\frac{\lambda {\erad}'}{2{\ephi}}\right)^{\!2}\right)\Bigg].
\end{align}
In this Hamiltonian constraint $\lambda$ (the constant polymerization parameter) appears explicitly and thus it is considered to be
{the} natural constant of the theory, while $r_0$
is a derived quantity that appears when integrating the equations of motion. 

However, because of the radiative properties of this black hole \cite{Alonso-Bardaji:2025qft},
it has been of recent interest to consider $r_0$, instead of $\lambda$, as the fundamental constant of the model.
In such a case, $r_0$ can be considered to be independent of $m$, and one can {automatically} apply the procedure of the main
text to construct the Hamiltonian {corresponding to the metric \eqref{eq.ourmetric}}.
More precisely, we see that this line element corresponds to the shape functions
$h_1=1$, $h_2=r$, and $h_3=1-r_0/r$ in \eqref{eq.modschw}. Replacing these in Eq.~\eqref{eq.Hwithh2r}, we see that the Hamiltonian constraint
can be expressed as {an} additive correction to the Hamiltonian of vacuum general relativity,
\begin{align}\label{ourh2}
    \ham_{r_0}:=\ham^{\rm GR}_{\rm vacuum}+2r_0\krad\kang.
\end{align}
This provides a clear example of how two seemingly distinct Hamiltonians [cf. \eqref{ourh1} and \eqref{ourh2}] can nevertheless define the same geometry \eqref{eq.ourmetric} through their dynamical flow in phase space. In itself, this is not surprising, as there is a lot of freedom in constructing the Hamiltonian constraint corresponding to a given geometry, as discussed in the main text. However, {as shown} in this particular case, we observe that the choice of which constant of the model {($\lambda$ or $r_0$)} is regarded as fundamental—and therefore allowed to appear explicitly in the Hamiltonian—plays a crucial role in determining which Hamiltonian constraint is the most natural one.

\section{Deformed Schwarzschild line element from phase space}\label{app.checkhs}

In this appendix, we will solve the system of equations \eqref{eq.dotexhs}--\eqref{eq.hamhsvanish} in the gauge $\erad=x^2$ and $\kang=0$. 
The conservation of the first gauge condition, $\dot{E}^x=0$ in \eqref{eq.dotexhs} fixes $N^x=0$. Then, the vanishing of the diffeomorphism constraint \eqref{eq.diffhsvanish} can be solved for $\krad=0$. Now, the vanishing of the Hamiltonian constraint \eqref{eq.hamhsvanish} reduces to 
\begin{align}
    0=\frac{4h_2^2\ephi'}{(\ephi)^2}-\frac{3}{\ephi}\frac{dh_2^2}{dx}+\frac{2\ephi}{h_2}\frac{d(h_1h_2)}{dx},
\end{align}  
which can be solved for 
\begin{align}
    \ephi=\varepsilon h_2\left(h_1-\frac{2m}{h_2}\right)^{-1/2},
\end{align}
with $m$ an integration constant and $\varepsilon=\pm1$. If we now turn to the conservation of the second gauge condition, $\dot{K}_\varphi=0$ in \eqref{eq.dotkphihs}, we find
\begin{align}
    0=\lapse\left(\frac{dh_1}{dx}+\left(\frac{h_1}{h_2}-\frac{h_2}{(\ephi)^2}\right)\frac{dh_2}{dx}\right)+\frac{2h_2^2}{(\ephi)^2}\frac{d\lapse}{dx},
\end{align}
and, enforcing the above conditions,
\begin{align}
    \lapse=c\left(h_1-\frac{2m}{h_2}\right)^{1/2},
\end{align}
with $c$ a nonvanishing constant that amounts for a trivial rescaling of the Killing field. We set it to one with no loss of generality. With this, all the equations \eqref{eq.dotexhs}--\eqref{eq.hamhsvanish} are satisfied. If we replace the solutions in the line element \eqref{eq.metrictx}, which now reads,
\begin{align}\label{eq.metrichs}
    ds^2= -N^2dt^2+\frac{(\ephi)^2}{h_2^2h_3}(dx+N^xdt)^2+\erad d\Omega^2,
\end{align}
we recover \eqref{eq.modschw} after relabeling $x$ as $r$.

\section{Checking covariance in the $\bar{\mu}$-loop black hole}\label{app.extra}

Let us point out that in Ref.~\cite{Kelly:2020uwj}, the metric \eqref{eq.mulqg} was derived as a solution of an effective
Hamiltonian{, which includes corrections motivated by loop quantum gravity,} in a certain {fixed} gauge.
However, the model is not covariant, and different gauge choices lead to different geometries. For instance, the polar gauge
produces the classical Schwarzschild geometry\footnote{It is straightforward to check that,
in the original reference \cite{Kelly:2020uwj}, {the functions} $b=0$, $N^x=0$, $E^b=x^{3/2}/(x-2 m)^{1/2}$, and $N=(1-2m/x)^{1/2}$ 
{solve} their equations of motion (4.7), (4.10), and (4.11), which, replacing in their line element (4.9), defines the exterior of the
Schwarzschild black hole.},
{which cannot be related to \eqref{eq.mulqg} by a coordinate transformation}.

{In this appendix} we explicitly show that \eqref{eq.mulqgham} is indeed {a covariant Hamiltonian}
associated with the geometry \eqref{eq.mulqg}. For that, we solve the {Hamilton equations
\begin{align}
\dot{E}^x=\{E^x,\int (N\ham_{\bar\mu}+N^x\diff)dx\},\\
\dot{K}_x=\{K_x,\int (N\ham_{\bar\mu}+N^x\diff)dx\},\\
\dot{E}^\varphi=\{E^\varphi,\int (N\ham_{\bar\mu}+N^x\diff)dx\},\\
\dot{K}_\varphi=\{K_\varphi,\int (N\ham_{\bar\mu}+N^x\diff)dx\},
\end{align}
along with the constraints $\diff=0$ and $\ham_{\bar\mu}=0$} 
in two different gauges, and then see that both solutions are related by a coordinate transformation.
First, in the Schwarzschild gauge that has been used to construct the Hamiltonian. Second, in the Gullstrand-Painlev\'e chart,
which is the line element appearing in the original Ref.~\cite{Kelly:2020uwj}. We recall that we are working with $r=\sqrt{\erad}$.

\subsection{Schwarzschild gauge}

We begin with the Schwarzschild gauge $\erad=x^2$ 
and $\kang=0$. The conservation of the first gauge condition,
\begin{align}
    0=\dot{E}^x=\shift\erad'+2\lapse\sqrt{\erad}\kang,
\end{align}
shows that we are defining a diagonal line element with $\shift=0$. The diffeomorphism constraint \eqref{eq.diff} sets $\krad=0$,
and  we can then {solve $\ham_{\bar\mu}=0$, as given by \eqref{eq.mulqgham},} for $\ephi=\pm{x^3}/{\sqrt{x^4-2m x^3+\alpha}}$, 
with $m$ a free integration constant. The conservation of the second gauge condition yields
\begin{align}
    0=\dot{K}_\varphi=\shift\kang'+\lapse'\frac{x^2}{(\ephi)^2}+\frac{\lapse}{2}\left(\frac{x}{(\ephi)^2}-\frac{x^4-3\alpha}{x^5}\right),
\end{align}
where, enforcing the solution for $\ephi$, we find $\lapse=c x/\ephi$, with $c$ a free integration constant that we can set to $c=\mathrm{sgn}(\ephi)$ with no loss of generality (it corresponds to a trivial rescaling of the time coordinate). 
Since $F_s=x^2$, replacing all the results in \eqref{eq.metrictx} {and relabeling $x$ as $r$}, the metric reads, as expected,
\begin{align}
    ds^2
    =-\left(1-\frac{2m}{r}+\frac{\alpha}{r^4}\right)dt^2+ \left(1-\frac{2m}{r}+\frac{\alpha}{r^4}\right)^{-1}dr^2+r^2d\Omega^2.
\end{align}
This result might seem trivial, as we are just undoing the process of Hamiltonian construction. Therefore, we now turn to the Gullstrand-Painlev\'e chart.

\subsection{Gullstrand-Painlev\'e gauge}

We {now begin with the gauge conditions} $\erad=x^2$ and $\ephi=x$, so that the spatial part of the metric is flat (recall that, in this case, $F_s=\erad$). 
Let us enforce the diffeomorphism constraint $\diff=0$ and the Hamiltonian constraint $\ham=0$,
\begin{align}
    0&=2\krad-\kang',\\
    0&=-\frac{1}{2}\left(\kang^2-\frac{3\alpha}{x^4}\right) -2x\krad\kang,
\end{align}
to find $\krad=\kang'/2$ and $\kang=\pm\frac{1}{x^2}\sqrt{R_S x^3-\alpha}$, with $R_S$ a free integration constant.
The conservation of the two gauge conditions yields
\begin{align}
0&=\dot{E}^x=\shift\erad'+2\lapse\sqrt{\erad}\kang=2x\left(\shift+\lapse\kang\right),\\
    0&=\dot{E}^\varphi=(\shift\ephi)'+\lapse\left(\frac{\ephi\kang}{{\sqrt{\erad}}}+2\sqrt{\erad}\krad\right)=(x\shift)'+\lapse(\kang+2x\krad),
\end{align}
and substituting the values {for $K_x$ and $K_\varphi$} obtained above in these expressions, we finally obtain
\begin{align}
    \lapse&= \varepsilon,\\
    \shift&=-\lapse\kang=\mp\frac{\varepsilon}{x^2}\sqrt{2m x^3 -\alpha},
\end{align}
with a trivial integration constant $\varepsilon=\pm1$. Finally, since $F_s=\erad=(\ephi)^2=x^2$,
in this gauge the metric \eqref{eq.metrictx} reads
\begin{align}
    ds^2
    =-\left(1-\frac{2m}{r}+\frac{\alpha}{r^4}\right)dt^2+2\varepsilon\sqrt{\frac{2m}{r}-\frac{\alpha}{r^4}}dtdr +dr^2+r^2d\Omega^2,
\end{align}
{after relabeling $x$ as $r$. 
This} is exactly the line {element  \eqref{eq.muoriginal} with $\varepsilon=1$ as given} in the original Ref.~\cite{Kelly:2020uwj}.

\bibliographystyle{bib-style}
\bibliography{biblio}

\providecommand{\noopsort}[1]{}\providecommand{\singleletter}[1]{#1}%
\providecommand{\href}[2]{#2}\begingroup\raggedright\begin{thebibliography}{10}

\bibitem{bardeen}
J.~Bardeen, ``Non-singular general relativistic gravitational collapse,'' Proceedings of the 5th International Conference on Gravitation and the Theory of Relativity, (Tiflis, U.S.S.R.), Tbilisi University Press (1968).

\bibitem{poisson:1988wc}
E.~Poisson and W.~Israel, ``{Structure of the Black Hole Nucleus},'' Class. Quant. Grav. {\bf 5} (1988) L201--L205.

\bibitem{Dymnikova:1992ux}
I.~Dymnikova, ``{Vacuum nonsingular black hole},'' Gen. Rel. Grav. {\bf 24} (1992) 235--242.

\bibitem{Hayward:2005gi}
S.~A. Hayward, ``{Formation and evaporation of regular black holes},'' Phys. Rev. Lett. {\bf 96} (2006) 031103, \href{http://arXiv.org/abs/gr-qc/0506126}{{\tt arXiv:gr-qc/0506126}}.

\bibitem{Mars:1996khm}
M.~Mars, M.~M. Mart{\'\i}n-Prats, and J.~M.~M. Senovilla, ``{Models of regular Schwarzschild black holes satisfying weak energy conditions},'' Class. Quant. Grav. {\bf 13} (1996), no.~5, L51--L58.

\bibitem{Borde:1996df}
A.~Borde, ``{Regular black holes and topology change},'' Phys. Rev. D {\bf 55} (1997) 7615--7617, \href{http://arXiv.org/abs/gr-qc/9612057}{{\tt arXiv:gr-qc/9612057}}.

\bibitem{Lemos:2011dq}
J.~P.~S. Lemos and V.~T. Zanchin, ``{Regular black holes: Electrically charged solutions, Reissner-Nordstr{\"o}m outside a de Sitter core},'' Phys. Rev. D {\bf 83} (2011) 124005, \href{http://arXiv.org/abs/1104.4790}{{\tt arXiv:1104.4790}}.

\bibitem{Bambi:2013ufa}
C.~Bambi and L.~Modesto, ``{Rotating regular black holes},'' Phys. Lett. B {\bf 721} (2013) 329--334, \href{http://arXiv.org/abs/1302.6075}{{\tt arXiv:1302.6075}}.

\bibitem{Frolov:2016pav}
V.~P. Frolov, ``{Notes on nonsingular models of black holes},'' Phys. Rev. D {\bf 94} (2016), no.~10, 104056, \href{http://arXiv.org/abs/1609.01758}{{\tt arXiv:1609.01758}}.

\bibitem{Ansoldi:2008jw}
S.~Ansoldi, ``{Spherical black holes with regular center: A Review of existing models including a recent realization with Gaussian sources},'' in {\em {Conference on Black Holes and Naked Singularities}}.
\newblock 2, 2008.
\newblock \href{http://arXiv.org/abs/0802.0330}{{\tt arXiv:0802.0330}}.

\bibitem{Simpson:2018tsi}
A.~Simpson and M.~Visser, ``{Black-bounce to traversable wormhole},'' JCAP {\bf 02} (2019) 042, \href{http://arXiv.org/abs/1812.07114}{{\tt arXiv:1812.07114}}.

\bibitem{Barcelo:2010vc}
C.~Barcel{\' o}, L.~J. Garay, and G.~Jannes, ``{Quantum Non-Gravity and Stellar Collapse},'' Found. Phys. {\bf 41} (2011) 1532--1541, \href{http://arXiv.org/abs/1002.4651}{{\tt arXiv:1002.4651}}.

\bibitem{Barcelo:2014cla}
C.~Barcel{\'o}, R.~Carballo-Rubio, L.~J. Garay, and G.~Jannes, ``{The lifetime problem of evaporating black holes: mutiny or resignation},'' Class. Quant. Grav. {\bf 32} (2015), no.~3, 035012, \href{http://arXiv.org/abs/1409.1501}{{\tt arXiv:1409.1501}}.

\bibitem{Barcelo:2015uff}
C.~Barcel{\'o}, R.~Carballo-Rubio, and L.~J. Garay, ``{Black holes turn white fast, otherwise stay black: no half measures},'' JHEP {\bf 01} (2016) 157, \href{http://arXiv.org/abs/1511.00633}{{\tt arXiv:1511.00633}}.

\bibitem{Modesto:2008im}
L.~Modesto, ``{Semiclassical loop quantum black hole},'' Int. J. Theor. Phys. {\bf 49} (2010) 1649--1683, \href{http://arXiv.org/abs/0811.2196}{{\tt arXiv:0811.2196}}.

\bibitem{Ashtekar:2018lag}
A.~Ashtekar, J.~Olmedo, and P.~Singh, ``{Quantum Transfiguration of Kruskal Black Holes},'' Phys. Rev. Lett. {\bf 121} (2018), no.~24, 241301, \href{http://arXiv.org/abs/1806.00648}{{\tt arXiv:1806.00648}}.

\bibitem{Kelly:2020uwj}
J.~G. Kelly, R.~Santacruz, and E.~Wilson-Ewing, ``{Effective loop quantum gravity framework for vacuum spherically symmetric spacetimes},'' Phys. Rev. D {\bf 102} (2020), no.~10, 106024, \href{http://arXiv.org/abs/2006.09302}{{\tt arXiv:2006.09302}}.

\bibitem{Alonso-Bardaji:2021yls}
A.~Alonso-Bardaji, D.~Brizuela, and R.~Vera, ``{An effective model for the quantum Schwarzschild black hole},'' Phys. Lett. B {\bf 829} (2022) 137075, \href{http://arXiv.org/abs/2112.12110}{{\tt arXiv:2112.12110}}.

\bibitem{Alonso-Bardaji:2022ear}
A.~Alonso-Bardaji, D.~Brizuela, and R.~Vera, ``{Nonsingular spherically symmetric black-hole model with holonomy corrections},'' Phys. Rev. D {\bf 106} (2022), no.~2, 024035, \href{http://arXiv.org/abs/2205.02098}{{\tt arXiv:2205.02098}}.

\bibitem{Belfaqih:2024vfk}
I.~H. Belfaqih, M.~Bojowald, S.~Brahma, and E.~I. Duque, ``{Black holes in effective loop quantum gravity: Covariant holonomy modifications},'' \href{http://arXiv.org/abs/2407.12087}{{\tt arXiv:2407.12087}}.

\bibitem{Rovelli:2014cta}
C.~Rovelli and F.~Vidotto, ``{Planck stars},'' Int. J. Mod. Phys. D {\bf 23} (2014), no.~12, 1442026, \href{http://arXiv.org/abs/1401.6562}{{\tt arXiv:1401.6562}}.

\bibitem{Boehmer:2007ket}
C.~G. Boehmer and K.~Vandersloot, ``{Loop Quantum Dynamics of the Schwarzschild Interior},'' Phys. Rev. D {\bf 76} (2007) 104030, \href{http://arXiv.org/abs/0709.2129}{{\tt arXiv:0709.2129}}.

\bibitem{Alesci:2020zfi}
E.~Alesci, S.~Bahrami, and D.~Pranzetti, ``{Asymptotically de Sitter universe inside a Schwarzschild black hole},'' Phys. Rev. D {\bf 102} (2020), no.~6, 066010, \href{http://arXiv.org/abs/2007.06664}{{\tt arXiv:2007.06664}}.

\bibitem{Han:2020uhb}
M.~Han and H.~Liu, ``{Improved effective dynamics of loop-quantum-gravity black hole and Nariai limit},'' Class. Quant. Grav. {\bf 39} (2022), no.~3, 035011, \href{http://arXiv.org/abs/2012.05729}{{\tt arXiv:2012.05729}}.

\bibitem{Alonso-Bardaji:2024tvp}
A.~Alonso-Bardaji, ``{Formation of nonsingular spherical black holes with holonomy corrections},'' Phys. Rev. D {\bf 111} (2025), no.~8, 084023, \href{http://arXiv.org/abs/2410.20529}{{\tt arXiv:2410.20529}}.

\bibitem{Ashtekar:2023cod}
A.~Ashtekar, J.~Olmedo, and P.~Singh, ``{Regular black holes from Loop Quantum Gravity},'' in {\em Regular Black Holes: Towards a New Paradigm of Gravitational Collapse}, C.~Bambi, ed.
\newblock Springer, Singapore, 2023.
\newblock \href{http://arXiv.org/abs/2301.01309}{{\tt arXiv:2301.01309}}.

\bibitem{Cano:2018aod}
P.~A. Cano, S.~Chimento, T.~Ort{\'\i}n, and A.~Ruip{\'e}rez, ``{Regular Stringy Black Holes?},'' Phys. Rev. D {\bf 99} (2019), no.~4, 046014, \href{http://arXiv.org/abs/1806.08377}{{\tt arXiv:1806.08377}}.

\bibitem{Brandenberger:2021jqs}
R.~Brandenberger, L.~Heisenberg, and J.~Robnik, ``{Through a black hole into a new universe},'' Int. J. Mod. Phys. D {\bf 30} (2021), no.~14, 2142001, \href{http://arXiv.org/abs/2105.07166}{{\tt arXiv:2105.07166}}.

\bibitem{Bonanno:2023rzk}
A.~Bonanno, D.~Malafarina, and A.~Panassiti, ``{Dust Collapse in Asymptotic Safety: A Path to Regular Black Holes},'' Phys. Rev. Lett. {\bf 132} (2024), no.~3, 031401, \href{http://arXiv.org/abs/2308.10890}{{\tt arXiv:2308.10890}}.

\bibitem{Lan:2023cvz}
C.~Lan, H.~Yang, Y.~Guo, and Y.-G. Miao, ``{Regular Black Holes: A Short Topic Review},'' Int. J. Theor. Phys. {\bf 62} (2023), no.~9, 202, \href{http://arXiv.org/abs/2303.11696}{{\tt arXiv:2303.11696}}.

\bibitem{Bambi:2023try}
C.~Bambi, ed., {\em {Regular Black Holes. Towards a New Paradigm of Gravitational Collapse}}.
\newblock Springer Series in Astrophysics and Cosmology. Springer, 2023.
\newblock \href{http://arXiv.org/abs/2307.13249}{{\tt arXiv:2307.13249}}.

\bibitem{Carballo-Rubio:2023mvr}
R.~Carballo-Rubio, F.~Di~Filippo, S.~Liberati, and M.~Visser, ``{Singularity-free gravitational collapse: From regular black holes to horizonless objects},'' in {\em Regular Black Holes: Towards a New Paradigm of Gravitational Collapse}, C.~Bambi, ed.
\newblock Springer, Singapore, 2023.
\newblock \href{http://arXiv.org/abs/2302.00028}{{\tt arXiv:2302.00028}}.

\bibitem{Carballo-Rubio:2025fnc}
R.~Carballo-Rubio {\em et al.}, ``{Towards a non-singular paradigm of black hole physics},'' JCAP {\bf 05} (2025) 003, \href{http://arXiv.org/abs/2501.05505}{{\tt arXiv:2501.05505}}.

\bibitem{Carballo-Rubio:2019fnb}
R.~Carballo-Rubio, F.~Di~Filippo, S.~Liberati, and M.~Visser, ``{Geodesically complete black holes},'' Phys. Rev. D {\bf 101} (2020) 084047, \href{http://arXiv.org/abs/1911.11200}{{\tt arXiv:1911.11200}}.

\bibitem{Bronnikov:2005gm}
K.~A. Bronnikov and J.~C. Fabris, ``{Regular phantom black holes},'' Phys. Rev. Lett. {\bf 96} (2006) 251101, \href{http://arXiv.org/abs/gr-qc/0511109}{{\tt arXiv:gr-qc/0511109}}.

\bibitem{Bronnikov:2012ch}
K.~A. Bronnikov, R.~A. Konoplya, and A.~Zhidenko, ``{Instabilities of wormholes and regular black holes supported by a phantom scalar field},'' Phys. Rev. D {\bf 86} (2012) 024028, \href{http://arXiv.org/abs/1205.2224}{{\tt arXiv:1205.2224}}.

\bibitem{Lavrelashvili:1992ia}
G.~V. Lavrelashvili and D.~Maison, ``{Regular and black hole solutions of Einstein Yang-Mills Dilaton theory},'' Nucl. Phys. B {\bf 410} (1993) 407--422.

\bibitem{Ovalle:2023ref}
J.~Ovalle, R.~Casadio, and A.~Giusti, ``{Regular hairy black holes through Minkowski deformation},'' Phys. Lett. B {\bf 844} (2023) 138085, \href{http://arXiv.org/abs/2304.03263}{{\tt arXiv:2304.03263}}.

\bibitem{Ayon-Beato:1998hmi}
E.~Ayon-Beato and A.~Garcia, ``{Regular black hole in general relativity coupled to nonlinear electrodynamics},'' Phys. Rev. Lett. {\bf 80} (1998) 5056--5059, \href{http://arXiv.org/abs/gr-qc/9911046}{{\tt arXiv:gr-qc/9911046}}.

\bibitem{Balart:2014cga}
L.~Balart and E.~C. Vagenas, ``{Regular black holes with a nonlinear electrodynamics source},'' Phys. Rev. D {\bf 90} (2014), no.~12, 124045, \href{http://arXiv.org/abs/1408.0306}{{\tt arXiv:1408.0306}}.

\bibitem{Ayon-Beato:2004ywd}
E.~Ayon-Beato and A.~Garcia, ``{Four parametric regular black hole solution},'' Gen. Rel. Grav. {\bf 37} (2005) 635, \href{http://arXiv.org/abs/hep-th/0403229}{{\tt arXiv:hep-th/0403229}}.

\bibitem{Dymnikova:2004zc}
I.~Dymnikova, ``{Regular electrically charged structures in nonlinear electrodynamics coupled to general relativity},'' Class. Quant. Grav. {\bf 21} (2004) 4417--4429, \href{http://arXiv.org/abs/gr-qc/0407072}{{\tt arXiv:gr-qc/0407072}}.

\bibitem{Bronnikov:2000yz}
K.~A. Bronnikov, ``{Comment on `Regular black hole in general relativity coupled to nonlinear electrodynamics'},'' Phys. Rev. Lett. {\bf 85} (2000) 4641.

\bibitem{Ayon-Beato:2000mjt}
E.~Ayon-Beato and A.~Garcia, ``{The Bardeen model as a nonlinear magnetic monopole},'' Phys. Lett. B {\bf 493} (2000) 149--152, \href{http://arXiv.org/abs/gr-qc/0009077}{{\tt arXiv:gr-qc/0009077}}.

\bibitem{DeFelice:2024seu}
A.~De~Felice and S.~Tsujikawa, ``{Instability of Nonsingular Black Holes in Nonlinear Electrodynamics},'' Phys. Rev. Lett. {\bf 134} (2025), no.~8, 081401, \href{http://arXiv.org/abs/2410.00314}{{\tt arXiv:2410.00314}}.

\bibitem{narlikar:1977nf}
J.~V. Narlikar and A.~K. Kembhavi, ``{Space-Time Singularities and Conformal Gravity},'' Lett. Nuovo Cim. {\bf 19} (1977) 517--520.

\bibitem{Bambi:2016wdn}
C.~Bambi, L.~Modesto, and L.~Rachwa{\l}, ``{Spacetime completeness of non-singular black holes in conformal gravity},'' JCAP {\bf 05} (2017) 003, \href{http://arXiv.org/abs/1611.00865}{{\tt arXiv:1611.00865}}.

\bibitem{Berej:2006cc}
W.~Berej, J.~Matyjasek, D.~Tryniecki, and M.~Woronowicz, ``{Regular black holes in quadratic gravity},'' Gen. Rel. Grav. {\bf 38} (2006) 885--906, \href{http://arXiv.org/abs/hep-th/0606185}{{\tt arXiv:hep-th/0606185}}.

\bibitem{Rodrigues:2015ayd}
M.~E. Rodrigues, E.~L.~B. Junior, G.~T. Marques, and V.~T. Zanchin, ``{Regular black holes in $f(R)$ gravity coupled to nonlinear electrodynamics},'' Phys. Rev. D {\bf 94} (2016), no.~2, 024062, \href{http://arXiv.org/abs/1511.00569}{{\tt arXiv:1511.00569}}. [Addendum: Phys.Rev.D 94, 049904 (2016)].

\bibitem{Junior:2015fya}
E.~L.~B. Junior, M.~E. Rodrigues, and M.~J.~S. Houndjo, ``{Regular black holes in $f(T)$ Gravity through a nonlinear electrodynamics source},'' JCAP {\bf 10} (2015) 060, \href{http://arXiv.org/abs/1503.07857}{{\tt arXiv:1503.07857}}.

\bibitem{DAmbrosio:2021zpm}
F.~D'Ambrosio, S.~D.~B. Fell, L.~Heisenberg, and S.~Kuhn, ``{Black holes in $f(Q)$ gravity},'' Phys. Rev. D {\bf 105} (2022), no.~2, 024042, \href{http://arXiv.org/abs/2109.03174}{{\tt arXiv:2109.03174}}.

\bibitem{Giacchini:2021pmr}
B.~L. Giacchini, T.~d.~P. Netto, and L.~Modesto, ``{Action principle selection of regular black holes},'' Phys. Rev. D {\bf 104} (2021), no.~8, 084072, \href{http://arXiv.org/abs/2105.00300}{{\tt arXiv:2105.00300}}.

\bibitem{Olmo:2022cui}
G.~J. Olmo and D.~Rubiera-Garcia, ``{Regular black holes in Palatini gravity},'' in {\em Regular Black Holes: Towards a New Paradigm of Gravitational Collapse}, C.~Bambi, ed.
\newblock Springer, Singapore, 2023.
\newblock \href{http://arXiv.org/abs/2209.05061}{{\tt arXiv:2209.05061}}.

\bibitem{Ziprick:2010vb}
J.~Ziprick and G.~Kunstatter, ``{Quantum Corrected Spherical Collapse: A Phenomenological Framework},'' Phys. Rev. D {\bf 82} (2010) 044031, \href{http://arXiv.org/abs/1004.0525}{{\tt arXiv:1004.0525}}.

\bibitem{Bueno:2024eig}
P.~Bueno, P.~A. Cano, R.~A. Hennigar, and {\'A}.~J. Murcia, ``{Dynamical Formation of Regular Black Holes},'' Phys. Rev. Lett. {\bf 134} (2025), no.~18, 181401, \href{http://arXiv.org/abs/2412.02742}{{\tt arXiv:2412.02742}}.

\bibitem{Carballo-Rubio:2025ntd}
R.~Carballo-Rubio, ``{Master field equations for spherically symmetric gravitational fields beyond general relativity},'' \href{http://arXiv.org/abs/2507.15920}{{\tt arXiv:2507.15920}}.

\bibitem{Alonso-Bardaji:2023vtl}
A.~Alonso-Bardaji and D.~Brizuela, ``{Spacetime geometry from canonical spherical gravity},'' Phys. Rev. D {\bf 109} (2024), no.~4, 044065, \href{http://arXiv.org/abs/2310.12951}{{\tt arXiv:2310.12951}}.

\bibitem{Bojowald:2023xat}
M.~Bojowald and E.~I. Duque, ``{Emergent modified gravity: Covariance regained},'' Phys. Rev. D {\bf 108} (2023), no.~8, 084066, \href{http://arXiv.org/abs/2310.06798}{{\tt arXiv:2310.06798}}.

\bibitem{Alonso-Bardaji:2025qft}
A.~Alonso-Bardaji, D.~Brizuela, and M.~Schneider, ``{Radiative properties of a nonsingular black hole: Hawking radiation and gray-body factor},'' JHEP {\bf 07} (2025) 189, \href{http://arXiv.org/abs/2504.13050}{{\tt arXiv:2504.13050}}.

\end{thebibliography}\endgroup

\end{document}